\documentclass[A4paper,prl,onecolumn,aps,floatfix,citeautoscript,nobibnotes,superscriptaddress,longbibliography]{revtex4-1}
\usepackage{xpatch} 
\makeatletter
\xpatchcmd{\@ssect@ltx}{\@xsect}{\protected@edef\@currentlabelname{#8}\@xsect}{}{}
\xpatchcmd{\@sect@ltx}{\@xsect}{\protected@edef\@currentlabelname{#8}\@xsect}{}{}
\makeatother

\usepackage{graphicx}
\usepackage{amsfonts,amsmath,amssymb,mathtools}
\usepackage{latexsym}
\usepackage{cases}
\usepackage{color}
\usepackage[colorlinks=true,allcolors=blue,breaklinks=true]{hyperref}
\usepackage[utf8]{inputenc}
\usepackage{siunitx}
\usepackage{orcidlink}
\usepackage{caption}

\AtBeginDocument{%
    \newwrite\bibnotes
    \def\bibnotesext{Notes.bib}
    \immediate\openout\bibnotes=\jobname\bibnotesext
    \immediate\write\bibnotes{@CONTROL{REVTEX41Control}}
    \immediate\write\bibnotes{@CONTROL{%
    apsrev41Control,author="08",editor="1",pages="1",title="0",year="1"}}
     \if@filesw
     \immediate\write\@auxout{\string\citation{apsrev41Control}}%
    \fi
}%

\def\change{\color{black}}

\newcommand*\sign{\mathop{}\!\mathrm{sgn}}
\newcommand*\diff{\mathop{}\!\mathrm{d}}

\begin{document}

\title{Boundary geometry controls a topological defect transition that determines lumen nucleation in embryonic development}

\author{Pamela C. Guruciaga\orcidlink{0000-0002-1399-6607}}
\thanks{\change These authors contributed equally to this work}
\email{pamela.guruciaga@embl.de}
\affiliation{Cell Biology and Biophysics Unit, European Molecular Biology Laboratory (EMBL), Meyerhofstra\ss e 1, 69117 Heidelberg, Germany}
\affiliation{Developmental Biology Unit, European Molecular Biology Laboratory (EMBL), Meyerhofstra\ss e 1, 69117 Heidelberg, Germany}

\author{Takafumi Ichikawa\orcidlink{0000-0002-0490-4176}}
\thanks{\change These authors contributed equally to this work}
\affiliation{Institute for the Advanced Study of Human Biology (WPI-ASHBi), Kyoto University, Kyoto, 606-8501, Japan}
\affiliation{Department of Developmental Biology, Graduate School of Medicine, Kyoto University, Kyoto, 606-8501, Japan}

\author{\change Steffen Plunder\orcidlink{0000-0002-3371-3790}}
\affiliation{Institute for the Advanced Study of Human Biology (WPI-ASHBi), Kyoto University, Kyoto, 606-8501, Japan}

\author{Takashi Hiiragi\orcidlink{0000-0003-4964-7203}}
\email{t.hiiragi@hubrecht.eu}
\affiliation{Institute for the Advanced Study of Human Biology (WPI-ASHBi), Kyoto University, Kyoto, 606-8501, Japan}
\affiliation{Department of Developmental Biology, Graduate School of Medicine, Kyoto University, Kyoto, 606-8501, Japan}
\affiliation{Hubrecht Institute, Uppsalalaan 8, 3584 CT Utrecht, The Netherlands}

\author{Anna Erzberger\orcidlink{0000-0002-2200-4665}}
\email{erzberge@embl.de}
\affiliation{Cell Biology and Biophysics Unit, European Molecular Biology Laboratory (EMBL), Meyerhofstra\ss e 1, 69117 Heidelberg, Germany}
\affiliation{Department of Physics and Astronomy, Heidelberg University, 69120 Heidelberg, Germany}

\date{Dated: April 18, 2025}

\begin{abstract}
        Topological defects determine the collective properties of anisotropic materials. How their configurations are controlled is not well understood however, especially in 3D. In living matter moreover, 2D defects have been linked to biological functions, but the role of 3D polar defects is unclear. Combining computational and experimental approaches, we investigate how confinement geometry controls surface-aligned polar fluids, and what biological role 3D polar defects play in tissues interacting with extracellular boundaries. We discover a charge-preserving transition between 3D defect configurations driven by boundary geometry and independent of material parameters, and show that defect positions predict the locations where fluid-filled lumina---structures essential for development---form within the confined polar tissue of the mouse embryo. Experimentally perturbing embryo shape beyond the transition point, we moreover create additional lumina at predicted defect locations. Our work reveals how boundary geometry controls polar defects, and how embryos use this mechanism for shape-dependent lumen formation. We expect this defect control principle to apply broadly to systems with orientational order.
\end{abstract}

\maketitle


Confinement plays a fundamental role in directing self-organisation~\cite{araujo2023steering} in various contexts including colloids~\cite{bradley2017janus}, developing tissues~\cite{trushko2020buckling,harmansa2023growth}, 
liquid crystals~\cite{lopez2011drops}, and crowds of people~\cite{sieben2017collective}. Controlling confining structures and predicting their impact on the collective properties of the confined bulk will reveal organisational principles of complex systems like living matter~\cite{barrat2023soft} and enable technological applications such as the design of 
optoelectronic~\cite{nys2020patterned} and microfluidic~\cite{daieff2020confined} devices.
Boundary effects are especially relevant for materials comprised of anisotropic particles, which possess orientational degrees of freedom and tend to align with their neighbours. 
In liquid crystals, for example, substrates with particular surface topography or chemical functionality~\cite{xin2023alignment} induce the alignment of adjacent molecules, creating an ordered layer that propagates into the bulk by elastic forces.
In this way, surface-induced alignment allows to achieve different desired configurations of the order parameter (OP) by tuning the boundary conditions, for example in nematics~\cite{sheng1976phase,sheng1982boundary,crawford1991surface}. However, the effects of boundary geometry on the bulk organisation of anisotropic materials is not well understood, particularly for curved surfaces in 3D.

Confinement also influences the internal organisation of living materials, e.g. how multicellular systems undergo morphogenesis in biofilms~\cite{nijjer2023biofilms} or during embryonic development~\cite{bondarenko2023embryo}. 
The boundary interactions between anisotropic living matter such as tissues consisting of polarised cells, and extracellular matrix (ECM) involve diverse biophysical processes, including adhesion and polarity signalling~\cite{o2001rac1,akhtar2013integrin,rasmussen2012laminin}, 
which affect ordering at the collective scale similar to boundary-induced alignment.
For instance, basement membrane layers promote specific orientations of adjacent cells for different types of polarity---parallel for epithelial planar ~\cite{davey2017planar,butler2017planar} and mesenchymal/migratory polarity~\cite{ladoux2016front,palmquist2022reciprocal,vicente2023order}, and perpendicular for apico-basal epithelial polarity~\cite{buckley2022apical}. Coherent apico-basal alignment is crucial for epithelial functions such as molecule transport across the tissue, and is linked to the formation of fluid-filled cavities (lumina)~\cite{buckley2022apical}. Orientational boundary effects could provide general mechanisms by which confining structures control multicellular organisation.

Many collective properties of ordered materials depend on the number and spatial distribution of topological defects, i.e.~points in the OP field where the orientation is not defined. 
In the case of nematic systems confined to a surface, geometric constraints have been shown to play an important role in determining defect configurations and dynamics~\cite{ienaga2023geometric,vafa2023periodic}.
In biological contexts, defects can guide diverse cellular processes~\cite{ardavseva2022topological} and have been shown to trigger the formation of structures such as fruiting bodies~\cite{copenhagen2021topological} and tentacles~\cite{maroudas2021topological}. However, despite their potential biological role, 3D polar defects and their relation to system geometry remain unexplored, with most investigations focusing on 2D domains and nematic systems, applicable in multicellular contexts where cells can be approximated as elongated units. These approaches have successfully predicted the collective dynamics of bacterial colonies~\cite{li2019data,yaman2019emergence,copenhagen2021topological}, cell populations~\cite{kawaguchi2017topological,ienaga2023geometric}, and epithelial sheets~\cite{saw2017topological,blanch2018turbulent}. However, in contexts where cellular polarity implies distinct sub-cellular regions with specific functions---such as {\change apico-basal polarity, in which the asymmetric organisation of organelles and cell-surface proteins enables directed transport~\cite{rodriguez2014organization}, or} mesenchymal {\change cell polarity driving active movement}~\cite{blanch2021integer,blanch2021quantifying}---, the polar nature of cells is relevant for understanding collective properties.

Here, we investigate how boundary effects lead to a surface--bulk coupling that permits controlling the properties of anisotropic materials through confining geometry, {\change focusing on boundaries capable of inducing bulk topological charge}. 
Specifically, we analyse how the mechanical properties and geometry of the boundaries {\change control} the ordering {\change and defect configurations} of a 3D polar fluid {\change confined by non-uniform surfaces. W}e identify three transitions: {\change two} in which defects in the bulk OP field appear out of a uniform field {\change driven by changes in boundary coupling and material response, and a third charge-preserving defect transition controlled by the geometry alone}.

We {\change apply these findings to} study the role of boundary geometry on the bulk organisation of {\change apico-basally} polarised cells, using the mouse epiblast---the tissue that forms the embryo proper during development---as an example. Orientational order of apico-basal polarity in the epiblast is linked to the formation of a central lumen, a critical event in normal mouse development~\cite{ichikawa2022ex,bedzhov2014self}. By associating the polar OP field with the local average cellular apico-basal polarity in the tissue, we find that 3D-field defect configurations are parameter-free predictors of lumen initiation sites. Moreover, using a recently developed \textit{ex vivo} culturing approach~\cite{ichikawa2022ex}, we experimentally perturb epiblast boundary shape to induce additional lumen nucleation sites at the predicted positions.

\section{Weak boundary-induced alignment in a 3D polar fluid}

{\change To investigate the mechanisms by which boundaries influence orientational order, we consider a} polar fluid {\change confined} in a 3D space $\Omega$ with volume $V_0$. {\change Such systems can be} characterised by a 3D vector OP $\mathbf p(\mathbf{r})$ 
{\change representing the average polarity of particles in a volume element~\cite{julicher2018hydrodynamic}, which allows to define the global degree of order
\begin{equation}\label{eq:globalOP}
    P\equiv \langle|\mathbf p|\rangle_{\Omega}=\frac{1}{V_0}\int_{\Omega}\diff V |\mathbf p| \,.
\end{equation}
When bulk and surface interactions are present and transients can be neglected, the local OP $\mathbf p(\mathbf{r})$ can be obtained by minimising the effective free energy functional
\begin{equation}\label{eq:F}
    \mathcal F[\mathbf p]
    =\int_{\Omega} \diff V f_{\mathrm B}(\mathbf p,\nabla\mathbf p) + \int_{\partial\Omega} \diff S f_{\mathrm S}(\mathbf p) \,,
\end{equation}
where $f_{\mathrm B}$ and $f_{\mathrm S}$ are the bulk and surface energy densities (see Supplementary Note).} 

Following the Landau--de Gennes approach to the nematic--isotropic transition in liquid crystals~\cite{de1971short}, we write the bulk energy density as $f_{\mathrm B} = f_{\mathrm R} + f_{\mathrm E}$.
To focus on boundary-induced order, we consider a restoring term $f_{\mathrm R}=a|\mathbf p|^2/2$ with $a\gtrsim 0$ favouring the disordered state $\mathbf p =0$.
The second term is given by the elastic contribution
\begin{equation}
    \label{eq:fE}
    f_{\mathrm E} =\frac{k_0}{2}\left(\nabla\cdot\mathbf p\right)^2 + \frac{k_1}{2}\left[\hat{\mathbf p}\cdot \left(\nabla\times\mathbf p\right)\right]^2 + \frac{k_2}{2}\left[\hat{\mathbf p}\times \left(\nabla\times\mathbf p\right)\right]^2
\end{equation}
with $\hat{\mathbf p}={\mathbf p}/|{\mathbf p}|$, 
in which we do not impose the one-constant approximation for the coefficients $k_0,\, k_1,\, k_2$ that penalise splay, twist, and bend distortions. Note that the coupling between the magnitude and orientational degrees of freedom in Eq.~\eqref{eq:fE} produces a regularizing cross-term absent in the classical Frank free energy {\change (see Supplementary Note)}.

The surface energy density $f_{\mathrm S}$ represents a weak anchoring interaction between the fluid and the confining surface $\partial\Omega$~\cite{prinsen2003shape,seyednejad2013confined,mertelj2017ferromagnetic}, where the OP at the boundary is not fixed as in the strong case~\cite{sheng1976phase,sheng1982boundary,crawford1991surface,prishchepa2005director}. Deviations from the \textit{preferred value} $\mathbf p_0$ at the surface are penalised as~\cite{ravnik2009landau,seyednejad2013confined}
\begin{equation}\label{eq:fS}
    f_{\mathrm S}=\frac{w}{2}\left(\mathbf{p}-\mathbf{p}_0\right)^2\,.
\end{equation}
The constant $w>0$ defines the anchoring length $\lambda\equiv w/a$.

Surface anchoring is known to control the defect structure in liquid crystals in the nematic phase~\cite{prishchepa2005director,lopez2011drops,tran2016lassoing}. 
In spherical confinement, a homeotropic {\change (i.e., perpendicular)} boundary condition produces a radial defect in the centre, while a {\change longitudinal} tangential alignment favours the appearance of two {\change surface radial defects known as \textit{boojums}} at the poles. 
For a polar OP, however, {\change the zoo of defects is larger due to the greater variety of possible orientations at the boundary}, and mixed boundary conditions---with both tangential and normal $\mathbf{p}_0$---remain unexplored {\change (Figs.~\ref{fig:defects_parameters}a-b)}, although a related problem was studied for liquid crystals in Ref.~\cite{prishchepa2005director}.
\begin{center}
    \includegraphics[scale=0.85]{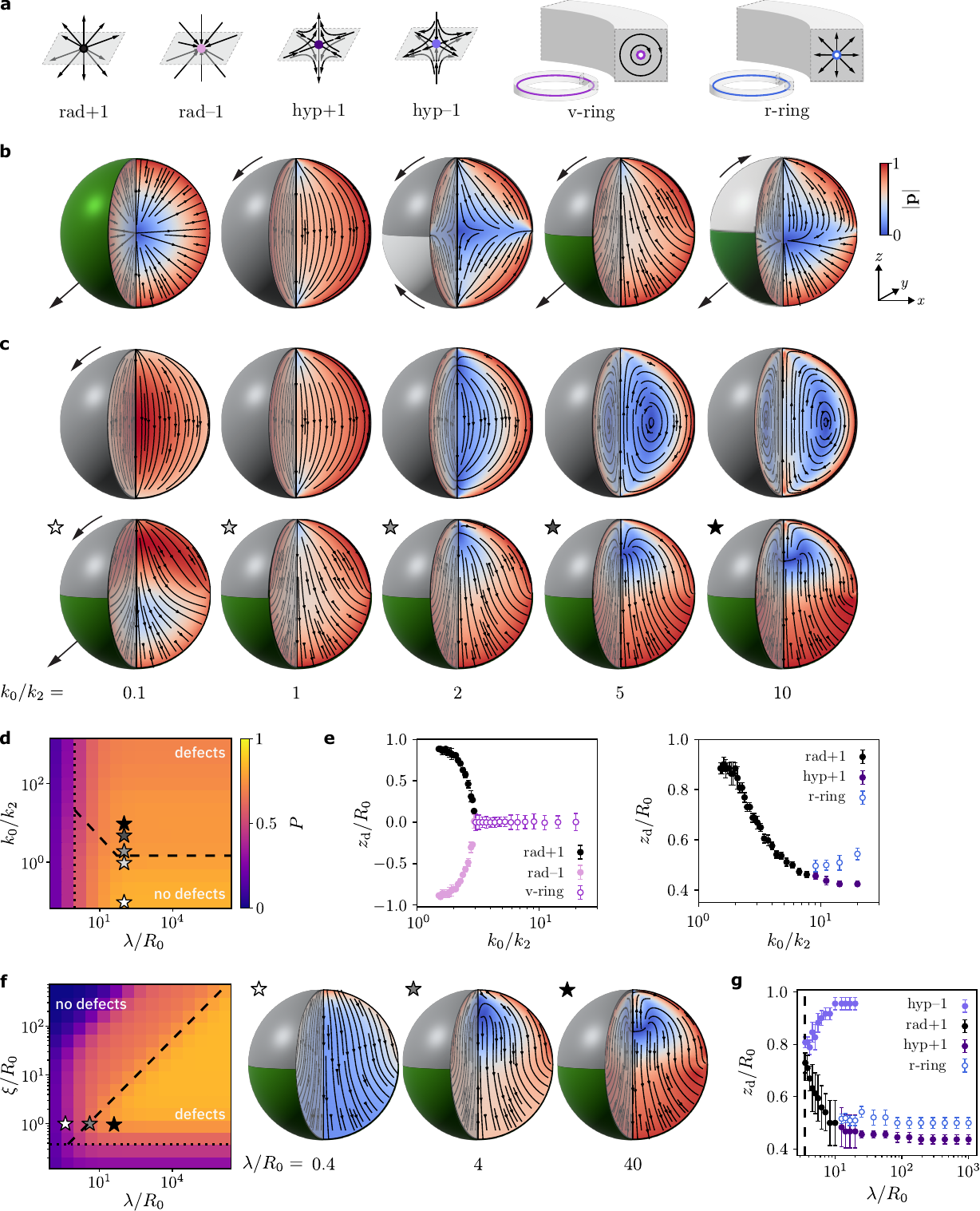}
    \captionof{figure}{\textbf{Material parameters determine the creation of 3D polar defects.}
        \textbf{a,}~Polar systems in 3D feature diverse defect structures, {\change shown schematically for the} radial $+1$ (black), radial $-1$ (pink), hyperbolic $+1$ (purple), hyperbolic $-1$ (lilac) hedgehogs, and the vortex (empty magenta) and radial (empty blue) disclination rings.
        {\change \textbf{b,} Streamlines and colour map show the local direction and magnitude of the equilibrium OP field in spherical confinement for anchoring and correlation lengths $\lambda/R_0=100$ and $\xi/R_0=1$ in the one-constant approximation ($k_0/k_2=1$) {\change with different boundary orientations} (from left to right: perpendicular, longitudinal parallel, parallel in the poles-to-equator direction, mixed, and inverted mixed). External arrows denote $\mathbf p_0$ on parallel-aligning (gray) and perpendicular-aligning (green) surfaces.}
        \textbf{c,}~OP fields with varying splay-to-bend ratio $k_0/k_2$ for {\change longitudinal} tangential (top) and mixed (bottom) boundaries ($\lambda/R_0=100$, $\xi/R_0=1$) {\change show how defects move from the boundary into the bulk as splay becomes the dominant mode}.
        {\change \textbf{d,}~The global degree of order $P$ [equation~\eqref{eq:globalOP}] for a sphere with mixed $\mathbf p_0$ as a function of the anchoring length $\lambda$ and the splay-to-bend ratio $k_0/k_2$ ($\xi/R_0=1$) shows that the system achieves global ordering (i.e., $P>0.5$) only if $\lambda>\lambda^*\approx R_0$ (dotted line). The dashed line separates defect-free and defect-containing regimes. Stars denote the parameter values of configurations in~\textbf{c}.}
        \textbf{e,}~{\change The defect height $z_{\rm{d}}$ shows how bulk defects enter from the boundary and change their position and configuration as the splay-to-bend ratio increases ($\lambda/R_0=100$, $\xi/R_0=1$).} The type and number of defects depends on the boundary preference ({\change left: longitudinal parallel, right: mixed}).
        \textbf{f,}~The global degree of order $P$ [equation~\eqref{eq:globalOP}] for a sphere with mixed $\mathbf p_0$ {\change in terms of the correlation and anchoring lengths, $\xi$ and $\lambda$ ($k_0/k_2 = 10$) shows that} the system remains disordered (i.e., $P<0.5$) even in the limit $\lambda\to\infty$ {\change for $\xi<\xi^*\approx 0.4R_0$ (dotted line)}. Stars denote the parameter values for example field configurations, where pairs of defects are created in order to accommodate the increasingly relevant boundary {\change preference}. The value $\zeta^*\approx 0.3R_0$ (dashed line) separates defect-free and defect-containing configurations. 
        \textbf{g,}~{\change The defect height $z_{\rm{d}}$ shows that the hyp$-1$ bulk defect moves up and becomes a hyperbolic boojum on the surface, while the rad$+1$ decomposes into a combination of a hyp$+1$ and a r-ring as the anchoring length $\lambda$ increases ($\xi/ R_0=1$, $k_0/k_2 = 10$).} 
        Points and error bars correspond to the mean and standard deviation of the position of the defect candidates (see~\nameref{sec:methods}). The vertical dashed line marks the value $\lambda^*=\xi^2/\zeta^*$.
    }
    \label{fig:defects_parameters}
\end{center}

\section{Material parameters determine the creation of 3D polar defects}

Mixed boundary conditions are relevant whenever a polar material contacts surfaces with different properties. 
To investigate how these non-uniform boundaries favouring distinct distortion modes affect bulk organisation, we analyse the impact of different elastic coefficients~\cite{revignas2020interplay,lavrentovich2024splay}.
Numerical minimisation of equation~\eqref{eq:F} in spherical confinement with tangential and mixed $\mathbf{p}_0$ reveals distinct field configurations and defect structures based on the splay-to-bend ratio $k_0/k_2$ (Fig.~{\change \ref{fig:defects_parameters}c}, see \nameref{sec:methods}). 
Note that the twist contribution vanishes {\change for an axially symmetric system}.
While there are only surface defects for $k_0/k_2 {\change \leq} 1$, bulk defects form and move away from the surface as the {\change splay-to-bend} ratio increases, {\change favouring bulk bend distortions over splay-inducing surface defects} (Figs.{\change \ref{fig:defects_parameters}c-e}), similar to the transition between nematic tactoids and toroids~\cite{lavrentovich2024splay}.
The type of bulk defects depends on the preferred orientation on the surface, with pole-to-pole tangential $\mathbf{p}_0$ giving rise in the $k_0\gg k_2$ limit to a toroidal vortex tube centred on the symmetry axis resembling the flow lines of a passive fluid in contact with an active surface~\cite{mietke2019minimal}. 
In contrast, the combination of tangential and normal preferred orientations results in a configuration with a hyperbolic point defect on the symmetry axis accompanied by a radial disclination ring around it.

{\change Within} the splay-dominated regime, the correlation length of the material {\change can be defined as} $\xi\equiv\sqrt{k_0/a}$. 
While the development of global order is promoted by the surface anchoring (and modulated by the surface--to--volume ratio, see Extended Data Fig.~\ref{ext:hemisphere}a), the correlation length determines the penetration of boundary-induced order into the bulk ($\xi>\xi^*\approx 0.4 R_0$ for a sphere, Fig.~\ref{fig:defects_parameters}f). The extrapolation length $\zeta\equiv k_0/w =\xi^2/\lambda$ sets the relative importance of bulk versus boundary interactions. Defects appear when surface anchoring dominates over the bulk $\zeta<\zeta^*\approx 0.3 R_0$, 
trading off the cost of bulk distortions in favour of better alignment with the preferred orientations at the boundaries, comparable to the transition in nematic tactoids~\cite{prinsen2003shape}. 
We find that the number and structure of defects depend on the confinement geometry. While in both spherical and hemispherical confinement a pair of radial (rad$+1$) and negative hyperbolic (hyp$-1$) hedgehogs nucleate at the transition, in the former the rad$+1$ defect further decomposes into a positive hyperbolic (hyp$+1$) hedgehog and radial disclination ring (r-ring) on increasing anchoring (Figs.~\ref{fig:defects_parameters}f-g, Extended Data Figs.~\ref{ext:hemisphere}a-b, {\change Supplementary Note}).
The topological charge for both geometries changes from neutral to $+1$ when the hyp$-1$ hedgehog moves up and becomes a hyperbolic boojum~\cite{volovik1983topological} on the cap surface (Fig.~\ref{fig:defects_parameters}g, Extended Data Fig.~\ref{ext:hemisphere}b). Note that the r-ring carries no charge~\cite{pollard2024morse}. 
In summary, we find that the competition between different distortion modes and boundary anchoring determines the global topological charge of boundary-aligned polar systems.

{\change 
\section{Material parameters of the mouse epiblast}

In living systems, boundary properties are often non-uniform. For example, the mouse epiblast at embryonic day 5.5 (E5.5) is confined by two distinct interfaces: an ECM layer, and a tissue--tissue interface with the extra-embryonic ectoderm (ExE; Figs.~\ref{fig:differentBCs_and_system}a-b). Epiblast cells develop apico-basal polarity, leading to directional cell--cell interactions via asymmetrically localised polarity and junctional protein complexes, which provide mechanical coupling and integration of biochemical signals~\cite{bedzhov2014self,ichikawa2022ex,mole2021integrin}. Interactions with ECM boundaries via e.g. basally localized integrin leads to the promotion of perpendicular apico-basal orientation via polarity signalling and mechanical feedback~\cite{ichikawa2022ex,mole2021integrin}, whereas cells orient predominantly parallel at the tissue-tissue interface.
\begin{figure}[thb]
    \centering
    \includegraphics[scale=0.85]{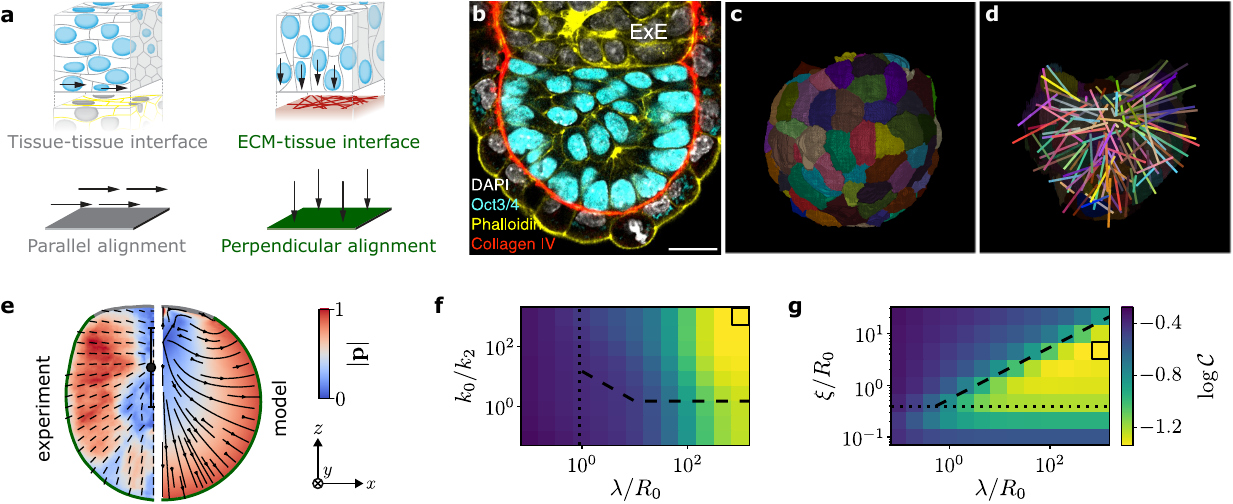}
    \caption{\textbf{\change Orientational interactions in the mouse epiblast.} 
    \textbf{a,}~Apico-basally polarised cells tend to align parallel to a tissue--tissue interface (left) and perpendicular to a tissue--ECM interface (right), similar to an anchoring effect [equation~\eqref{eq:fS}].
    \textbf{b,}~Immunofluorescence image stained for an ECM component (Collagen IV), actin (Phalloidin), an epiblast marker (Oct3/4), and DNA (DAPI) shows the central cross-section of a representative mouse embryo after \textit{ex vivo} culture for 18 hours from E4.5. At this stage, the epiblast consists of apico-basally polarised cells (cyan) in contact with a cup-shaped ECM layer (red) and a tissue--tissue interface with the ExE. Scale bar \SI{20}{\micro\meter}. Adapted from Ref.~\cite{ichikawa2022ex}.
    {\change 
        \textbf{c-d,}~Based on membrane segmentations of the individual epiblast cells from 3D images~(\textbf{c}), cell long axes~(\textbf{d}), representing each cell's unsigned orientation, were extracted for $n=5$ embryos at E5.25. Scale bar \SI{20}{\micro\meter}.
        \textbf{e,}~We obtain an estimate of the experimental director field and strength of alignment (left; lines and colour map, respectively) by computing the average cell orientations weighted by the corresponding cell elongation. Data also shown in Ref.~\cite{ichikawa2025boundary}. Using only the field magnitude, we fit the model parameters and find the polar OP that best represents the data (right; $\xi/R_0=4.6$, $\lambda/R_0=10^4$, $k_0/k_2=10^3$). Both fields present a topological defect (black point in the experimental case, with error bars denoting the standard deviation of the position).
        \textbf{f-g,}~Evaluating the cost function $\mathcal{C}$ (see \nameref{sec:methods}) over a range of splay-to-bend ratios $k_0/k_2$, anchoring lengths $\lambda$, and correlation lengths $\xi$ reveals that the system is robustly represented by a splay-dominated~(\textbf{f}), boundary-aligned~(\textbf{g}) polar material, in the defect-containing regime.
    }
    }
   \label{fig:differentBCs_and_system}
\end{figure}



To assess how such cell--cell and cell--boundary interactions influence the average cellular orientations, we performed 3D-gel embedded embryo culture, a technique that recapitulates \textit{in utero} embryonic development while allowing for \textit{in toto} monitoring, measurement, and manipulation~\cite{ichikawa2022ex}. Using inverted light-sheet microscopy, we imaged embryos with fluorescently labelled membranes and nuclei, observing frequent neighbour exchanges and fluid-like tissue dynamics over the course of this developmental stage (Supplementary Videos~\ref{vid:epiblast_dynamics_20} and~\ref{vid:epiblast_dynamics_5}). Segmenting the epiblast cells, we quantified their principal axes of inertia to determine their elongation and orientation within the tissue (Figs.~\ref{fig:differentBCs_and_system}c-d, see \nameref{sec:methods}). Taking advantage of the approximate axial symmetry of epiblasts at this stage, we averaged these measurements across rotational slices, and subsequently across the ensemble of five embryos, to calculate the characteristic nematic OP field (Fig.~\ref{fig:differentBCs_and_system}e). Because the underlying observables are invariant under polarity inversion, they cannot yield a direct estimate of the apico-basal polarity field: the nematic tensor captures the cell's orientation coherence, but not the polarity direction. However, the magnitude of the nematic field can provide an unsigned readout of the polarity magnitude, assuming that local inversions of apico-basal orientation are sufficiently rare to be neglected. 
Using this measured nematic magnitude, we estimated the relative strength of cell--boundary and cell--cell interaction modes by fitting the material parameters in equations~\eqref{eq:fE} and~\eqref{eq:fS} to the data (see \nameref{sec:methods}). 
We find that the epiblast tissue is best represented by a correlation length $\xi/R_0=4.6$, an anchoring length $\lambda/R_0=10^4$ and an splay-to-bend ratio $k_0/k_2=10^3$, situating the embryo within the splay-dominated, defect-containing regime (Figs.~\ref{fig:differentBCs_and_system}f-g). The direction of the OP field $\mathbf{p}$ obtained with these values resembles well the experimental nematic director field, even though the latter was not directly included in the fitting process. In particular, both fields present a radial defect on the symmetry axis (Fig.~\ref{fig:differentBCs_and_system}e). The good agreement between the confined polar fluid theory and the experimental measurements suggests that cellular configurations in the epiblast are indeed predominantly controlled by orientational cell--cell and cell--boundary interactions. In an accompanying study, we systematically analysed cell alignment patterns throughout the peri- to early post-implantation developmental stages, and found an increase in the anchoring and correlation length parameters during development, in line with an observed enrichment in integrin and ECM components at the epiblast boundary~\cite{ichikawa2025boundary}.
}

\section{Boundary shape controls charge-preserving defect transitions}

\begin{figure*}[htb]
    \centering
    \includegraphics[scale=0.85]{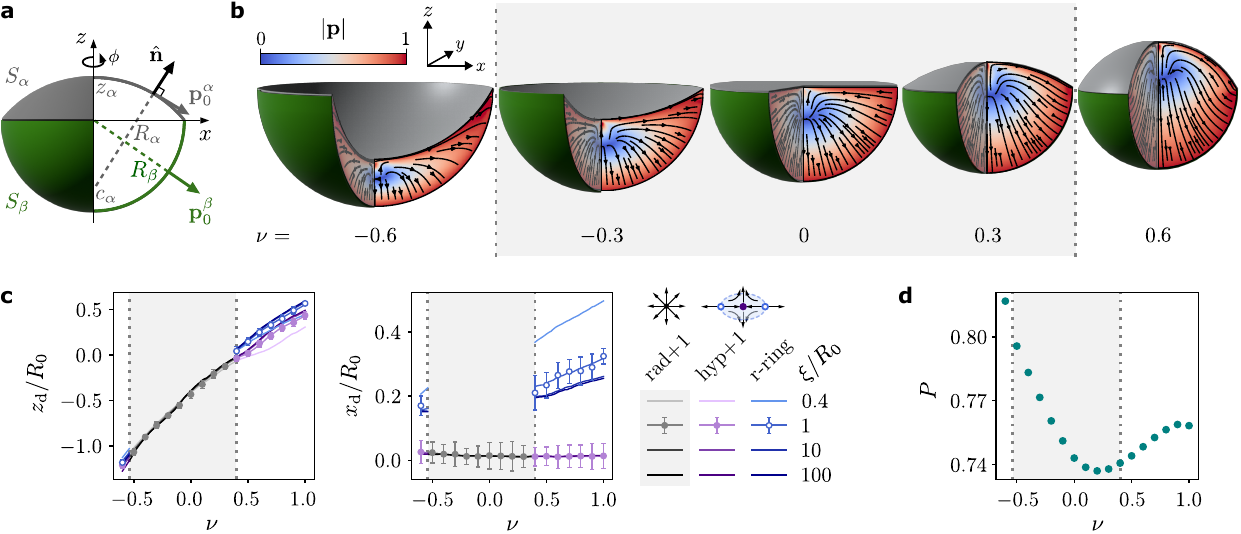}
    \caption{\textbf{Confining geometry controls charge-preserving transitions between different defect configurations.}
        {\change \textbf{a,}~We parametrise the class of acorn shapes with $\nu\equiv{z_{\alpha}}/{R_{\beta}}$. The base $S_{\beta}$ is defined as a hemisphere with radius $R_{\beta}$ centred at the origin of coordinates, while the spherical cap $S_{\alpha}$ has radius $R_{\alpha}$, centre $(0, c_{\alpha})$ on the symmetry axis, and maximum height $z_{\alpha}$. Vectors $\mathbf{p}_0^{\alpha}$ and $\mathbf{p}_0^{\beta}$ depict the mixed preferred orientation on the boundary.}
        {\change\textbf{b,}}~The OP fields for different {\change values of the acorn parameter} $\nu$ at fixed {\change anchoring and correlation} lengths $\lambda/ R_0=10^6$ and $\xi/ R_0=1$ (i.e., in the highly ordered, defect-containing regime) with {\change splay-to-bend ratio} $k_0/k_2 = 10$ show transitions between a radial hedgehog, and a hyp$+1$ defect surrounded by a radial disclination ring. 
        {\change\textbf{c,}}~{\change Height $z_{\mathrm{d}}$ and distance to the symmetry axis $x_{\mathrm{d}}$ (equivalent to the radius, in the case of the r-ring) as functions of the acorn parameter $\nu$.} The transitions {\change between defect configurations} occur when the shape parameter exceeds the critical values $\nu_{\mathrm c}^+=\change{0.38\pm 0.04}$ and $\nu_{\mathrm c}^-=\change{-0.56\pm 0.09}$ (dotted lines). The transition points are invariant to changes in the material parameters, and defect positions undergo little variation (see also Extended Data Fig.~\ref{ext:nucrit}). Points and error bars correspond to the mean and standard deviation of the position of the defect candidates (see~\nameref{sec:methods}).
        {\change\textbf{d,}}~The global degree of order $P$ also depends on {\change the acorn parameter} $\nu$, but the system remains always in the highly ordered ($P>0.5$) regime ($\lambda/ R_0=10^6$, $\xi/ R_0=1$, $k_0/k_2 = 10$).
    }
    \label{fig:defects_geometry}
\end{figure*}

Motivated by the shape of the mouse epiblast, we systematically examine the role of boundary geometry on 3D polar OP field configurations in the family of axis-symmetric \textit{{\change ideal} acorn} shapes, whose geometry is fully parametrised by $\nu\equiv{z_{\alpha}}/{R_{\beta}}$ with $z_{\alpha}$ the maximum height of the spherical-shell cap and ${R_{\beta}}$ the radius of the hemispherical base (Fig.~\ref{fig:defects_geometry}a, see \nameref{sec:methods}). While $\nu=0$ corresponds to a flat tissue--tissue interface, $0<\nu\leq 1$ (respectively $-1<\nu<0$) represent convex (respectively concave) interfaces, with $\nu=1$ the spherical case. The total volume $V_0$ is fixed and defines a length scale $R_0\equiv\left(3V_0/4\pi\right)^{1/3}$.

Fixing the material length scales in the splay- and surface-dominated regime, we vary the acorn parameter $\nu$ (Fig.~\ref{fig:defects_geometry}b-c), and find that the defect configuration undergoes a transition from radial hedgehog to a combination of hyp$+1$ point defect plus r-ring at {\change critical values} $\nu_{\mathrm c}^+$ and $\nu_{\mathrm c}^-$ (Fig.~\ref{fig:defects_geometry}b-c, {\change see also Extended Data Fig.~\ref{ext:flippedmixed}}). 
Notably, the surface--to--volume ratio and the global degree of order $P$ (Fig.~\ref{fig:defects_geometry}d) differ distinctly at these {\change two points}. 
Measuring the position of the defects (Fig.~\ref{fig:defects_geometry}c) shows that their height grows approximately linearly with the acorn parameter, and that the r-ring is nucleated with a finite radius. This renders the transitions discontinuous, comparable to the first-order transition in {\change nematics} with strong homeotropic boundary conditions~\cite{mkaddem2000fine}. 
Minimisations over intervals {\change of the anchoring and correlation lengths and the splay-to-bend ratio} $\lambda/R_0\in\left[10^2,10^6\right]$, $\xi/R_0\in\left[\xi^*, \sqrt{\lambda\zeta^*}\,\right)/R_0$ {\change and $k_0/k_2\in[10,10^2$]} reveal that the transition points and the position of the defects are robust to changes in material properties (Fig.~\ref{fig:defects_geometry}c and Extended Data Fig.~\ref{ext:nucrit}): the transition from the radial hedgehog to the hyp$+1$ and r-ring defects occurs independently of the material parameters at $\nu_{\mathrm c}^+=\change{0.38\pm 0.04}$ and $\nu_{\mathrm c}^-=\change{-0.56\pm 0.09}$, and the r-ring radius changes only $2$-fold over an almost three orders of magnitude variation in {\change the correlation length}.
In summary, we report a geometry-controlled transition in the defect structure of a boundary-aligned polar fluid, and show that defect locations are robust to changes in the material {\change parameters}. 

\section{Polar defects determine sites of lumen nucleation}

\begin{figure*}[htb]
    \centering
    \includegraphics[scale=0.85]{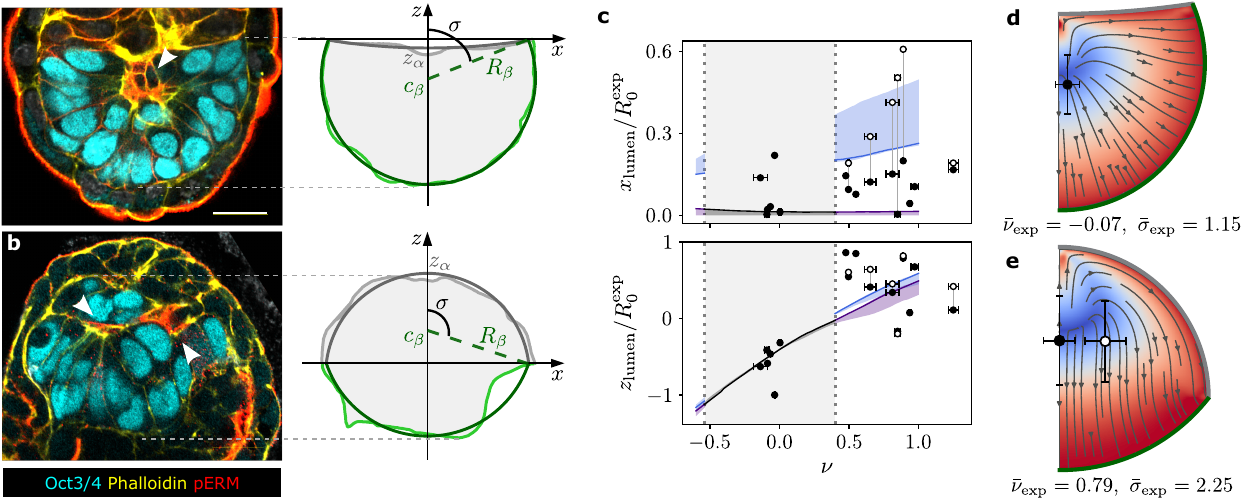}
    \caption{\textbf{Defect positions correspond to sites of lumen nucleation in the mouse epiblast.}
        \textbf{a-b,}~Representative immunostaining images show the central cross-section of a control (\textbf{a}) and a shape-manipulated (\textbf{b}) mouse embryo developed \textit{ex vivo} for 18 hours from E4.5, stained for an apical surface marker (pERM), actin (Phalloidin), and an epiblast marker (Oct3/4).
        Lumen initiation sites are highlighted with white arrowheads. Scale bars \SI{20}{\micro\meter}. To quantify the boundary shapes, we fit segmentations (light green: tissue--ECM boundary, light gray: tissue--tissue boundary) with a combination of two spherical caps, where the base is centered at $(0,c_{\beta})$. This generalised acorn family is parameterised by $\sigma\equiv\cos^{-1}\left(-c_{\beta}/R_{\beta}\right)$ and $\nu\equiv z_{\alpha}/R_{\beta}$.
        \textbf{c,}~Lumen centroid positions plotted against individually fitted $\nu_{\mathrm{exp}}$ show that control embryos ($n=6$) with shapes $\nu\in (\nu_{\mathrm{c}}^-,\nu_{\mathrm{c}}^+)$ (shaded region between dotted lines) contain single lumina, whereas $6$ out of $10$ embryos with shapes perturbed beyond the transition point present additional sites (empty points; vertical lines connect lumina from the same embryo). Coloured shaded regions show the defect position ranges for ideal acorns with {\change anchoring and correlation lengths} $\lambda/R_0=10^6$ and $\xi/R_0\in[0.4,100]$ and {\change splay-to-bend ratio} $k_0/k_2 = 10$ as in Fig.~\ref{fig:defects_geometry}c. {\change Solid lines predict---without further fitting---lumen nucleation positions for the epiblast correlation and anchoring length estimates.} The volume $V_0^{\mathrm{exp}}$ of the epiblast tissue defines $R_0^{\mathrm{exp}}=\left(3V_0^{\mathrm{exp}}/4\pi\right)^{1/3}$ for each embryo.
        Error bars are determined by residual bootstrapping of the epiblast boundaries (see \nameref{sec:methods}).
        \textbf{d-e,}~{\change Slice of OP fields calculated in the average shape of control (\textbf{d}) and manipulated (\textbf{e}) epiblasts {\change for the material parameter estimates}. 
        The average lumen nucleation sites ({\change full and empty black points}) are close to where the {\change corresponding} OP field presents defects. Error bars denote the standard deviation.}
        }
    \label{fig:lumina}
\end{figure*}

Topological defects are known to affect macroscopic properties of the systems they inhabit~\cite{ardavseva2022topological}. In living systems, 2D nematic defects have been shown to function as organising centres for biological processes like cell extrusion~\cite{saw2017topological} or out-of-plane bending during morphogenesis~\cite{hoffmann2022theory}. However, how 3D polar defects influence the multicellular organisation {\change of apico-basally polarised cells} remains unexplored. 
{\change In the epiblast,} we find that small fluid-filled cavities tend to form between the cells near positions where {\change both the experimentally estimated orientation field, and the theoretical} OP field of a correspondingly confined polar system has a radial hedgehog defect (Figs.~{\change\ref{fig:differentBCs_and_system}e}, \ref{fig:defects_geometry}b and~\ref{fig:lumina}a, {\change Supplementary Videos~\ref{vid:epiblast_dynamics_20} and~\ref{vid:epiblast_dynamics_5}}). {\change Radial positive and hyperbolic defects in the continuum orientation field mark positions where contacts between multiple apical surfaces are likely.} Indeed, such lumina typically nucleate between the apical surfaces of cells, which feature distinct molecular compositions and functional properties, with vesicle trafficking or active osmolyte pumps localising asymmetrically along the apico-basal axis~\cite{paul2003na+,rathbun2020cytokinetic,bryant2010molecular,sigurbjornsdottir2014molecular}. {\change Trans-interactions between neighbouring cells through junctional complexes localised near the apical domains are moreover required to seal and eventually inflate the cavity~\cite{pasquier2024inverted}, and actomyosin contractility at the apical surface can drive the constriction of multicellular rosettes~\cite{guyomar2024asymmetry}.}

{\change For a confining surface with uniform perpendicular preferred orientation, the continuum theory predicts the robust formation of a single lumen initiation site at the centre of the tissue (Fig.~\ref{fig:defects_parameters}b). To test this prediction, we experimentally perturbed the boundary organisation of the epiblast by immunosurgically removing precursors of the ExE tissue, thereby obtaining a near-spherical epiblast tissue surrounded everywhere by an ECM-rich boundary (Extended Data Figs.~\ref{ext:ExE-devoid}a-k). Indeed, we find that ExE-devoid epiblasts robustly present a rosette structure with a lumen in the centre (Extended Data Fig.~\ref{ext:ExE-devoid}l). This observation is in line with the typical behaviour of pluripotent stem cell systems, which self-organise into apico-basally polarised structures which robustly form a single central lumen when embedded in Matrigel, a condition that mimics uniform ECM confinement~\cite{bedzhov2014self,shahbazi2017pluripotent,rosa2024protocol}.}

{\change Moreover,} the geometry-driven transition we identified suggests that tissue shape controls the distribution of lumen nucleation sites, implying that externally manipulating boundary geometry may permit inducing additional lumina. 
We expect in particular a difference in the spatial distribution and number of lumen nucleation sites between epiblasts with shapes falling within and beyond the range $\nu\in (\nu_{\mathrm c}^-,\nu_{\mathrm c}^+)$, with the former containing sites only close to the central symmetry axis, and the latter presenting an additional distribution at a distance similar to the radius of the disclination ring.

To test these predictions,
we experimentally perturbed the shape of the epiblast, specifically the tissue--tissue interface. The curvature of this interface depends on a preceding morphogenetic event, the inward folding of the adjacent extra-embryonic ectoderm~\cite{ichikawa2022ex,christodoulou2019morphogenesis}. Disrupting this process results in embryos with deformed interfaces between the epiblast and this extra-embryonic tissue.
We obtained mouse embryos cultured for 18 hours from E4.5, of which epiblasts presented various shapes resulting from natural variability in the control condition (Fig.~\ref{fig:lumina}a) or experimental manipulation (Fig.~\ref{fig:lumina}b) by blocking the formation of the extra-embryonic ectoderm (see \nameref{sec:methods}). Note that the epiblast tissue volume {\change remained unchanged by this perturbation} [$\bar{V}_0^{\mathrm{exp}} \approx (2.3 \pm 0.4)\times10^{5}\,\SI{}{\micro\meter\cubed}$, see \nameref{sec:methods}].
To {\change better capture} the shape variations of these embryos, we consider a two-parameter family of shapes {\change defined by the acorn parameter $\nu$ and a new parameter $\sigma$, which reflects the vertical displacement of the centre of the base $S_{\beta}$ from the origin} (Figs.~\ref{fig:lumina}a-b).
Quantifying the shapes of the 16 embryos, we obtained {\change $\nu_{\mathrm{\exp}}\in[-0.14,\,0]$ and $\sigma_{\mathrm{\exp}}\in[0.8,\,1.43]$} for control embryos ($n = 6$; Fig.~\ref{fig:lumina}a, Extended Data Fig.~\ref{ext:sigma}), and {\change $\nu_{\mathrm{\exp}}\in[0.48,\,1.25]$ and $\sigma_{\mathrm{\exp}}\in[1.42,\,2.82]$} for embryos with a morphogenetic perturbation ($n = 10$; Fig.~\ref{fig:lumina}b, Extended Data Fig.~\ref{ext:sigma}). We excluded from this analysis embryos which did not undergo elongation of the whole embryo along the distal-proximal axis or did not reach the epiblast cell number of 50~\cite{ichikawa2022ex}. We also discarded one embryo in which the epiblast--ECM interface was concave and hence incompatible with our family of shapes.
We estimated the positions of lumen initiation sites using immunostaining against apical surface components and identified single sites near the central axis in embryos with $\nu_{\mathrm{c}}^-<\nu<\nu_{\mathrm{c}}^+$, and additional sites---as predicted---for the majority of embryos with $\nu>\nu_{\mathrm{c}}^+$ (Fig.~\ref{fig:lumina}c). Importantly, {\change the defect positions in the OP field for the average shape of control and manipulated embryos predict the average nucleation positions in each group quantitatively, without parameter fitting (Figs.~\ref{fig:lumina}d-e). This agreement is particularly striking given the biological complexity of the system, and the relatively small number of cells involved: despite variability in the single-cell configurations, the coarse-grained field correctly anticipates the average sites for rosette/lumen formation, suggesting that the underlying defect structure is a robust feature linking tissue geometry to polarity organisation.}
Taken together, these results show that lumina nucleate near the polar defect positions {\change of the continuum theory}, and that external manipulation of embryo shape induces the creation of additional nucleation sites near the position predicted by the geometry-driven defect transition we identified.

\section{Discussion and outlook} 

Confinement and boundary effects impact the bulk organisation of complex materials. We show how interactions with confining surfaces drives the creation of different topological defects in a polar fluid, and identify distinct roles for the confining geometry and bulk and surface mechanical parameters: while the competition between bulk distortion and surface anchoring governs the transitions between defect-free and defect-containing states, the \emph{types} of defects observed and the transitions between different defect configurations depend solely on the geometry. Moreover, in the surface-dominated regime, 3D defect positions are robust against anchoring and correlation length variations. We thus report a novel, charge-preserving defect transition controlled by confining geometry. {\change Indeed, because the preferred orientation $\mathbf{p}_0$ is defined within the surface, the boundary geometry affects how this field projects into the 3D bulk. As a result, shape changes can modulate internal orientations without requiring changes in material parameters.} 

In living systems, the impact of topological defects on macroscopic spatial organisation is not limited to distortion effects; they can also localise regulatory molecules that drive subsequent biological processes~\cite{ardavseva2022topological}. Examples include the activation of cell--death signals leading to cell extrusion~\cite{saw2017topological}, or assembling organisational centres during morphogenesis~\cite{hoffmann2022theory,guillamat2022integer}.  
For the mouse epiblast---a tissue consisting of apico-basally polarised cells---we find that defects guide the formation of fluid-filled cavities. Although lumina fulfil critical biological functions in tissue organisation and cell fate specification~\cite{durdu2014luminal,ryan2019lumen,kim2021deciphering}, how their spatial arrangement within a tissue arises {\change has been} unclear. The molecular processes that drive lumen {\change initiation} in the epiblast as well as in numerous other tissues~\cite{sigurbjornsdottir2014molecular}, such as active osmolyte pumping, localise to structures at the apical side of cells~\cite{paul2003na+,rathbun2020cytokinetic}. {\change Moreover, lumen initiation requires trans-interactions between the apical-domain proteins of neighbouring cells, for example between the junctional complexes required to seal and eventually inflate the cavity. Topological defects {\change in the continuum field} at which the tails of the polarity arrows converge mark spatial locations where contacts between the apical surfaces of multiple cells are overall most likely to form. Despite the relatively small number of cells in the mouse epiblast, and the overall variability in the specific configurations of the cells across different embryos,} we indeed found that defect positions are quantitative parameter-free predictors of lumen nucleation sites in this system. Furthermore, we accomplished inducing additional lumen nucleation sites by experimentally perturbing the boundary geometry beyond the transition point, where the correspondingly confined polar fluid presents an additional defect.
That the spatial distribution of early luminal structures follows the predicted geometry-dependent changes suggests that embryo shape regulates lumen formation via orientational effects. Lumen formation might therefore serve as a shape-sensing checkpoint mechanism in the embryo, determining successful amniotic cavity development.

It will be exciting to investigate the interplay between surface-induced alignment and boundary geometry in other experimental systems to challenge the generality of this mechanism for the spatial organisation of defect-associated structures. Importantly, since defect configurations depend almost exclusively on the geometry, the relevant observables are directly accessible from imaging data. Quantitative predictions can be made without fitting of material parameters, indeed without having to make detailed assumptions about specific bulk or surface properties. This robustness not only renders experimental testing feasible, but also suggests that spatial control of defect-associated processes via boundary geometry is a general phenomenon across systems with orientational degrees of freedom, independent of scale or material specifics. 

For multicellular systems, our formalism offers a general and tractable framework for studying cell--ECM interactions. 
Using a single anchoring parameter to capture the coupling between cellular polarity and extra-cellular materials like basement membranes will allow the investigation of the feedback between geometry changes and boundary-induced order from active ECM remodelling by cellular processes---properties distinguishing multicellular systems from passive anisotropic materials in confinement.

In conclusion, our investigations reveal how boundary geometry controls defect configurations within polar materials, and uncover a new biological function for 3D polar defects in multicellular systems. More broadly, that shape can control defect configurations independently of material specifics constitutes a general geometry-dependent organisational principle. Having demonstrated its predictive power in a system as complex and intractable as a living embryo, we expect this defect-control mechanism to be relevant in diverse orientational systems.




\clearpage

\section{Methods}
\label{sec:methods}


\subsection{Geometrical definitions}

\subsubsection{Ideal acorn}

If the total volume $V_0$ of the system is fixed, geometrical parameters $c_{\alpha}$, $R_{\alpha}$ and $R_{\beta}$ (see Fig.~\ref{fig:differentBCs_and_system}d in the main text) are defined solely by $\nu\equiv z_{\alpha}/R_{\beta}$ through the system of equations
\begin{equation}
    \begin{cases}
        R_{\alpha}^2 = R_{\beta}^2+c_{\alpha}^2\\
        \nu R_{\beta } =c_{\alpha} +\sign\,{\nu}\, R_{\alpha} \\
        V_0 = \dfrac{2\pi}{3} R_{\beta}^3 + \sign\,{\nu}\,\dfrac{\pi}{3} \left( R_{\alpha}-|c_{\alpha}| \right)^2 \left( 2R_{\alpha}+|c_{\alpha}| \right)
    \end{cases}
    .
\end{equation}

The cap $S_{\alpha}$---representing the tissue--tissue interface---can be parameterised in spherical coordinates as $r=r_{\alpha}(\theta) =
c_{\alpha}\cos\theta+\sqrt{R_{\alpha}^2-c_{\alpha}^2\sin^2\theta}$ for $\theta\in[0,\pi/2]$ (respectively $\theta\in[\pi/2,\pi]$) if $\nu>0$ (respectively $\nu<0$).
Note that this includes $\nu=1$, since in that case $c_{\alpha}=0$ and $R_{\alpha}=R_{\beta}$. If $\nu=0$, however, $S_{\alpha}$ is given by the disk with $r\in[0,R_{\beta}]$ and $\theta=\pi/2$. Finally, the base $S_{\beta}$---corresponding to the tissue--ECM interface---is simply $r=R_{\beta}$ and $\theta\in[\pi/2,\pi]$ for all $\nu$. In all cases, $\phi\in[0,2\pi)$.
With these definitions, the mixed preferred orientation in the spherical basis consists of $\mathbf p_0^{\beta}=(1,0,0)\ \forall\nu$ on $S_{\beta}$, and
\begin{widetext}
    \begin{equation}
        \mathbf p_0^{\alpha} =
        \begin{cases}
            (1,0,0) &\quad\text{if}\quad \nu=0 \\
            \left( \sin\theta - \sign\,\nu\,g_{\alpha}(\theta) \sin\theta\cos\theta,\ \cos\theta + \sign\,\nu\,g_{\alpha}(\theta) \sin^2\theta,\ 0 \right) &\quad\text{if}\quad 0<|\nu|<1 \\
            (0,1,0) &\quad\text{if}\quad \nu=1 
        \end{cases}
    \end{equation}
\end{widetext}
with 
\begin{equation}\label{eq:galpha}
    g_{\alpha}(\theta)=\frac{r_{\alpha}(\theta)}{\sqrt{R_{\alpha}^2-r_{\alpha}^2(\theta)\sin^2\theta}}
\end{equation}
on $S_{\alpha}$.
Note that the tissue--tissue interaction could in principle be modelled as nematic. However, we expect no significant difference, since the outward-normal $\mathbf p_0^{\beta}$ imposed by the tissue--ECM boundary and the polar character of the constituents break this symmetry and favour the north-south orientation given by $\mathbf p_0^{\alpha}$.

\subsubsection{Generalised acorn}

In a more general scenario, the base $S_{\beta}$ of the acorn can be centred at a point $(0,\ c_{\beta})$ different from the origin of coordinates. Such a system can be characterised by two dimensionless parameters: $\nu$, as introduced before, and $\sigma\equiv\cos^{-1}\left(-c_{\beta}/R_{\beta}\right)$ (see Figs.~\ref{fig:lumina}a-b in the main text). 
Considering a constant volume $V_0$, the equations that geometrical parameters $c_{\alpha}$, $R_{\alpha}$, $c_{\beta}$ and $R_{\beta}$ must satisfy are
\begin{equation}
    \begin{cases}
        R_{\alpha}^2 = R_{\beta}^2+(c_{\beta}-c_{\alpha})^2+2(c_{\beta}-c_{\alpha})R_{\beta}\cos\sigma\\
        \nu R_{\beta } =c_{\alpha} +\sign\,{\nu}\, R_{\alpha} \\
        V_0 = \dfrac{\pi}{3}R_{\beta}^3\left(2-\cos\sigma\right)\left(1+\cos\sigma\right)^2 + \nu \dfrac{\pi}{6} R_{\beta}^3 \left( 3\sin^2\sigma + \nu^2 \right)
    \end{cases}
    \ \ \ .
\end{equation}

While $S_{\alpha}$ remains the same as in the ideal case, $S_{\beta}$ is now too a spherical cap with $r=r_{\beta}(\theta) = c_{\beta}\cos\theta+\sqrt{R_{\beta}^2-c_{\beta}^2\sin^2\theta}$. By construction, the polar angle is still $\theta\in[\pi/2,\pi]$, and $\phi\in[0,2\pi)$ as always.
The normal to $S_{\beta}$ is no longer a constant and is written $\mathbf p_0^{\beta}= \left( -\cos\theta +g_{\beta}(\theta) \sin^2\theta,\ \sin\theta + g_{\beta}(\theta) \sin\theta\cos\theta,\ 0 \right)$ with $g_{\beta}(\theta)$ defined analogously to equation~\eqref{eq:galpha}.

\subsection{Numerical details}

We used the FEniCSx library DOLFINx~\cite{ScroggsEtal2022} to implement the finite-element method in Python3. We minimised equation~\eqref{eq:F} in a 3D, symmetry-agnostic parametrisation for selected relevant parameters and observed that no spontaneous breaking of axial symmetry occurred {\change (see Supplementary Note)}. Thereafter, the mesh (with resolution $0.05$) was parameterised in terms of $r$ and $\theta$ only, corresponding to a constant-$\phi$ slice. 
We computed the variation of $\mathcal{F}$ [equation~\eqref{eq:F}] with respect to the 3D vector OP $\mathbf{p}$ in the direction of a test function $\varphi$ to derive its weak formulation. The resulting non-linear problem was solved using a Newton solver with a relative tolerance of $10^{-6}$.

Number and type of defects in the OP field were determined by visual inspection of the phase and direction of $\mathbf{p}$.
Utilizing that $|\mathbf{p}|\to0$ in the vicinity of defects due to the large local distortions they induce, we infered the position of a defect as the central point of a mesh cell where $|\mathbf{p}|=0$, surrounded by a region with $|\mathbf{p}|>0$. The numerical uncertainty of these estimates thus depended on the size of the disordered region surrounding each defect and the resolution of the mesh. We defined a threshold $p^*$ and checked in each cell whether $|\mathbf{p}|<p^*$. The value of $p^*$ was fixed for each simulation, though we varied it conveniently from one to another in order to pre-select the least possible number of candidate points per defect (but in all cases, $p^*\leq 10^{-2}$). When more than one candidate points were selected, the location of the defect was calculated as the average of the candidate points. Error bars were determined as the standard deviation of their spatial distribution.


\subsection{Culture and imaging of peri-implantation mouse embryos}

To experimentally {\change characterize epiblast cells and} manipulate the tissue--tissue boundary in the peri-implantation mouse embryos, 3D-gel embedded embryo culture {\change (3D-geec)} was performed as described in Ref.~\cite{ichikawa2022ex}. 
In brief, mouse embryos at E4.5 were recovered from dissected uteri and cultured in a mixture of Matrigel--collagen I. Mural trophectoderm (mTE) was microsurgically removed to enable polar trophectoderm (pTE) invagination in the control condition, whereas maintaining mTE intact blocked pTE invagination and generated a boundary perturbation {\change in 18 hours of culture}.

{\change
3D-geec embryos were live-imaged using an inverted light-sheet microscope (Bruker, Luxendo, InVi SPIM), as described previously~\cite{ichikawa2022ex}. Briefly, embryos were embedded in a $15 \,\SI{}{\micro\liter}$ gel mix within the V-shaped sample holder attached with transparent FEP foil, carefully positioned so that they are in proximity but do not attach to the foil that would disrupt morphogenesis via adhesion. After gel solidification, embryos were immersed in $75 \,\SI{}{\micro\liter}$ IVC1 medium and further covered with $200 \,\SI{}{\micro\liter}$ mineral oil to prevent evaporation. The sample holder was mounted in an environmentally controlled incubation box with 5\% CO2 and 5\% O2 at $37^{\circ}$C. 
InVi SPIM was equipped with a Nikon 25x/1.1NA water immersion detective objective and a Nikon 10x/0.3 NA water immersion illumination objective. The illumination plane and focal plane were aligned before the imaging session and maintained during the imaging. Images were taken every 20 minutes by a CMOS camera (Hamamatsu, ORCA Flash4.0 V2) with line-scan mode in LuxControl (Luxendo). The imaged volume was $425.98\times425.98\times400 \,\SI{}{\micro\meter\cubed}$ with a physical voxel size of $0.208 \times 0.208 \times 1.000 \,\SI{}{\micro\meter\cubed}$. The lasers and filters used were $488\,\SI{}{\nano\meter}$ and BP525/50, and $561\,\SI{}{\nano\meter}$ and LP561 to image GFP and tdTomato fluorophores, respectively. Exposure time for each plane was set to $50\,\SI{}{\milli\second}$.

To remove polar TE and following ExE, mouse embryos recovered at E3.5 were subjected to immunosurgery, as described previously~\cite{ichikawa2022ex}. E3.5 blastocysts were recovered by flushing dissected oviducts and uteri with global® medium with HEPES (LifeGlobal, LGGH-050). Zona pellucida (ZP) was removed from blastocysts with pronase (0.5\% w/v Protease, Sigma P8811, in global® medium containing HEPES supplemented with 0.5\% Polyvinylpyrrolidone, Sigma, P0930) treatment for 2-3 minutes at $37^{\circ}$C. Blastocysts were washed in \SI{10}{\milli\liter} droplets of global® medium (LifeGlobal, LGGG-050) and further incubated in serum containing anti-mouse antibody (Cedarlane, CL2301, Lot no. 049M4847V) diluted 1:3 with global® medium for 30 minutes at $37^{\circ}$C. Following brief washes in global® medium with HEPES three times, embryos were incubated in guinea pig complement (Sigma, S1639, Lot no. SLBX9353) diluted with global® medium in a 1:3 ratio for 30 minutes at $37^{\circ}$C. Lysed cells and remaining debris were removed by gentle pipetting with a narrow glass capillary. The isolated inner cell masses (ICMs) were cultured in \SI{10}{\milli\liter} drops of global® medium in a petri dish (Falcon, 351008) covered with mineral oil (Sigma, M8410) and incubated at $37^{\circ}$C with 5\% CO2 for 24 hours. Once the PrE cells covered the epiblast cells, ICMs were cultured in 3D-geec for 24 or 48 hours, corresponding to Day2 and Day3 samples.
}

Embryos were fixed after {\change 3D-geec} culture with 4\% paraformaldehyde (FUJIFILM Wako, 166-23251) in PBS for 30 minutes and subsequently permeabilised with 0.5\% Triton X-100 (Nacalai, 12967-32) in PBS for 30 minutes. Embryos were incubated in 3\% BSA (Sigma, A9647) and 0.05\% Triton X-100 in PBS overnight at $4^{\circ}$C and then subjected to immunostaining. Primary antibodies against Oct3/4 (Santa Cruz Biotechnology, sc-5279 AF647), {\change Gata4 (R\&D systems, AF2606),} Collagen IV (Millipore, AB756P), and pERM (Cell Signaling, 3726) were diluted at 1:100. Donkey anti-rabbit IgG Alexa Fluor Plus 488 (Invitrogen, A32790), {\change Donkey anti-goat IgG Alexa Fluor Plus 488 (Invitrogen, A32814),} DAPI (Invitrogen, D3571), and Alexa Fluor Plus 555 Phalloidin (Invitrogen, A30106) were simultaneously used at the secondary antibody staining.

Images were obtained by LSM880 or LSM980 equipped with a C-Apochromat 40x/1.2 NA water immersion objective (Zeiss).

\subsection{Image analysis}

The epiblast tissue marked by Oct3/4 positive cells {\change and the inside lumen were} manually segmented with Napari~\cite{napari2023}. Labels for the entire area were drawn every 30 slices with \SI{0.16}{\micro\meter} interval, followed by the plugin ``napari-label-interpolator'' to fill the entire volume. Then the label statistics function in the plugin ``napari-simpleitk-image-processing'' was used to count the number of voxels of the epiblast {\change and lumen} volume.
We checked that perturbation of the tissue--tissue boundary introduced only geometric changes that preserved total volume. Indeed, average epiblast tissue volume was $(1.8 \pm 0.4)\times10^{5}\,\SI{}{\micro\meter\cubed}$ in the manipulated cohort and $(2.3 \pm 0.4)\times10^{5}\,\SI{}{\micro\meter\cubed}$ in the control condition.
{\change For the nascent nucleated lumina, their} position measurement was performed based on the space encircled with pERM signals with Fiji~\cite{schindelin2012fiji}, using the middle cross-section, which was extracted by the 3D rotation function in Imaris (Bitplane).  

{\change
Pre-processing for machine-learning-based membrane segmentation was performed with Fiji~\cite{schindelin2012fiji}. Membrane segmentation and custom model training were performed using Cellpose~\cite{stringer2021cellpose,pachitariu2022cellpose} GUI or CLI using a bash script. Manual correction of segmentation and cell tracking were performed with Napari~\cite{napari2023} (Supplementary Video~\ref{vid:3D_stack}). 
Segmentation to process the 3D images of the membrane signal was carried out as described in Ref.~\cite{ichikawa2025boundary}. 
The details of the individual steps are described below.
\begin{enumerate}
    \item \textit{Pre-processing}: Airyscan images were binned to $0.1647 \times 0.1647 \times 0.1600 \,\SI{}{\micro\meter\cubed}$ by averaging $2 \times 2 \times 1$ voxels, followed by isotropic transformations of voxels to a cube with a length of $0.1632 \pm 0.0002 \,\SI{}{\micro\meter}$. To quantify spatial parameters, we set the length of the voxel to $0.1632 \,\SI{}{\micro\meter}$ and ignored the associated error. Two channels for E-cadherin and phalloidin signals of the isotropically scaled Airyscan images were combined to generate a ubiquitous membrane channel by summing their signal intensities. The combined membrane channels were used as inputs for segmentation.
    \item \textit{Developing custom segmentation models}: Since the pre-trained models provided by Cellpose 2.0 required an insuperable amount of manual correction for our 3D images, we developed custom segmentation models tailored for Airyscan images. In brief, a pre-processed 3D image of an E5.0 stage embryo was resliced into three orthogonal planes (XY, YZ, and XZ), and five slices were extracted in each plane at a 50-slice interval across the entire image. The resulting slices were subjected to 2D segmentation using the Cellpose pre-trained model, CPx, followed by manual corrections of all cells in the slice using Napari. Following the instructions, two embryo datasets, each consisting of 15 pairs of slices and ground truth masks, were used to train the neural network, generating ``AS\_model\_1''. To further improve AS\_model\_1, we conducted additional training by incorporating two extra embryo datasets into the neural network, resulting in ``AS\_model\_2''. While both models showed comparable performance on images that have high signal-to-noise ratios, AS\_model\_2 is more robust to the images with various signal-to-noise ratios. Therefore, we used AS\_model\_2 in this study. 
    \item \textit{Batch segmentation}: For 3D segmentation of Airyscan images, segmentation parameters, such as flow\_threshold, cellprob\_threshold, and stitch\_threshold, were set to default values, except for the cell diameter, which was set to 60 pixels. The segmentation model ``AS\_model\_2'' was chosen for this process.
    \item \textit{Manual correction}: Segmentation errors were manually corrected by referring to the original images, and epiblast cells were selected and counted in Napari using a custom plugin (napari-labelselector, unpublished). The corrected segmentation masks were saved as `.tif' files for further analysis.
\end{enumerate}
}

{\change

\subsection{Determination of the experimental nematic field}

\subsubsection{Epiblast cell principal inertia vectors}
We first convert the epiblast segmentation into a mesh representation using the Python package scikit-image~\cite{scikit-image}. Each segmented cell is reconstructed as an individual mesh, and a global mesh for the entire epiblast region is generated. To ensure consistency in the analysis, all meshes are post-processed to correct face orientations, and only the largest connected component is retained. The final epiblast mesh stores cell labels as vertex attributes, enabling cell-specific calculations. We then use the trimesh library~\cite{trimesh} to compute various geometric and topological properties, including cell volume, surface area, centroid, and principal inertia vectors. Additionally, the centroid of the entire epiblast region is determined to facilitate further spatial analyses.

\subsubsection{Distal--proximal axis}
Imaging data is annotated with at least six points in Napari~\cite{napari2023}, through which a regression plane approximately separates the boundary between the tissue--tissue and tissue--ECM boundary regions. One point is annotated to denote the distal tip of the epiblast. We define the rotation axis as a line that passes through the epiblast centroid and the annotated tip point. Two points are also annotated to indicate the transverse edge of the epiblast, the mean position of which is used to establish a plane and define angle $\phi=0$.

\subsubsection{Weighted nematic average over rotational slices}
To construct a nematic field representing average cell order in a given epiblast, we first reorient the coordinate system such that the $z$-axis aligns with the distal-proximal axis, previously annotated. The spatial coordinates are then rescaled to normalise the epiblast volume to unity. At each spatial location $\mathbf{r}=(x,y,z)$, we extract the major principal inertia vector $v(\mathbf{r})$ and its corresponding inertia components $\lambda_1\leq\lambda_2\leq\lambda_3$ from the occupying cell mesh. The elongation axis is then defined as $v'(\mathbf{r})=\eta(\mathbf{r})v(\mathbf{r})$, where the shape anisotropy factor $\eta(\mathbf{r})\equiv\left(\lambda_3-\sqrt{\lambda_1\lambda_2}\right)/\lambda_3$ ranges from $0$ for spheres up to $1$ for infinitely elongated needles. Next, we define $M = 36$ equidistant angles $\phi_m$, generating rotational slices around the $z$-axis. For each slice, we use the rotation matrix $R(\phi_m)$ which maps from the rotational slice to the $xz$-plane to obtain elongation axes within the $xz$-plane as $v''_m(\mathbf{r})=R(\phi_m)v'\big(R^\intercal(\phi_m)\mathbf{r}\big)$. We then compute the weighted Landau-de Gennes $Q$-tensor, defined as 
\begin{equation}\label{eq:Q-tensor}
    Q(\mathbf{r})\equiv\frac{1}{M}\sum_{m=1}^M ||v''_m(\mathbf{r})|| \left( \frac{d}{d-1}\frac{v''_m(\mathbf{r})\otimes v''_m(\mathbf{r})}{||v''_m(\mathbf{r})||^2} - \frac{I}{d-1} \right)
\end{equation}
for $d=3$.

\subsubsection{Cross-embryo averaged nematic field}
We calculate the embryo-averaged $Q$-tensor $Q=\sum_{k=1}^K Q_k/K$, where $K$ is the number of embryos and $Q_k$ denotes the previously computed $Q$-tensors of individual embryos [equation~\eqref{eq:Q-tensor}]. 
We then take its principal eigenvector $V(\mathbf{r})$ and the corresponding eigenvalue $\Lambda(\mathbf{r})$ as the (unsigned) director field and the strength of the nematic alignment, respectively. 
For visualisation in 2D, we evaluate $V(\mathbf{r})$ over a uniform grid in the $xz$-plane, displaying only the in-plane components $(V_x,V_z)$, and use $\Lambda(\mathbf{r})$ as a heat map. The latter is subsequently used as experimental data for fitting the theoretical model, as described below. The position of the defect in the director field is estimated as the average of the points where the alignment strength $\Lambda(\mathbf{r})$ is below a threshold value of $0.05$, while its type is determined by visual inspection of the direction of $V(\mathbf{r})$.

\subsubsection{Cross-embryo averaged boundary curve}
For each embryo, we compute the intersection of the epiblast mesh with the $M$ rotational slice planes defined by angles $\phi_m$. These boundary curves are converted into polar coordinates $(r,\theta)$, with the epiblast centroid as the origin. The curves are resampled at equidistant angles, and their radial distances are averaged across all slices to produce a smooth, rotationally averaged boundary.
We then compute the cross-embryo average boundary by averaging with respect to all rotational slices from multiple embryos. 
Finally, the averaged boundary is transformed back into Cartesian coordinates for visualisation in the $xz$-plane.

}

\subsection{Shape model fitting}
\label{sec:shape_model_fitting}

Given the collections of points $\left\{\left(x^{\mathrm{exp}}_{\mu i},\, 0,\, z^{\mathrm{exp}}_{\mu i}\right)\right\}_{i=1,\ldots,N_{\mu}}$ corresponding to the segmented tissue--tissue ($\mu=\alpha$) and tissue--ECM ($\mu=\beta$) contours, we identify the centroid of the epiblast 
and the two points $\mathbf b_{\mathrm{left}}$, $\mathbf b_{\mathrm{right}}$ where the different boundaries meet. We define the axis of symmetry as the line passing through the centroid and perpendicular to the line between $\mathbf b_{\mathrm{left}}$ and $\mathbf b_{\mathrm{right}}$; the point where these lines intersect defines the origin of coordinates. 
After referring the contour points to this origin, we transform them to spherical coordinates, $\left\{\left(r^{\mathrm{exp}}_{\mu i},\, \theta^{\mathrm{exp}}_{\mu i},\, 0\right)\right\}_{i=1,\ldots,N_{\mu}}= \left\{\mathbf{r}^{\mathrm{exp}}_{\mu i}\right\}_{i=1,\ldots,N_{\mu}} = \mathbf{r}^{\mathrm{exp}}_{\mu}$, and minimise the cost function 
\begin{equation}
    \label{eq:costfunction}
    h\left(\mathbf{r}^{\mathrm{exp}}_{\alpha}, \mathbf{r}^{\mathrm{exp}}_{\beta}; \kappa_{\alpha}, {\change \gamma_{\alpha}}, \kappa_{\beta}, {\change \gamma_{\beta}}\right) = \sum_{{\mu}=\alpha,\beta}\,
    \sum_{i=1}^{N_{\mu}} \left[\mathcal{R}\left(\mathbf{r}^{\mathrm{exp}}_{\mu i}; \kappa_{\mu}, {\change \gamma_{\mu}} \right)\right]^2 \,.
\end{equation}
Here, 
\begin{equation}
    \label{eq:residuals}
    \mathcal{R}\left(\mathbf{r}^{\mathrm{exp}}_{\mu i}; \kappa_{\mu}, {\change \gamma_{\mu}} \right) \equiv r^{\mathrm{exp}}_{\mu i} - r(\theta^{\mathrm{exp}}_{\mu i}; \kappa_{\mu}, {\change \gamma_{\mu}})
\end{equation}
is the residual between the radius of experimental point $i$ of boundary $\mu$ and the fitting function
\begin{equation}
    r(\theta; \kappa_{\mu}, {\change \gamma_{\mu}}) = \frac{\cos\theta}{\change \gamma_{\mu}}+\sqrt{\frac{1}{\kappa_{\mu}^2}-\frac{\sin^2\theta}{\change \gamma_{\mu}^2}}
\end{equation}
evaluated at that point, with $\kappa_{\mu}\equiv 1/R_{\mu}$ (i.e., the curvature of the spherical cap) and {\change 
$\gamma_{\mu}\equiv 1/c_{\mu}$}. {\change This parametrisation, albeit counter-intuitive, allows us to easily fit flat caps, for which both the radius $R_{\mu}$ and the height $c_{\mu}$ of the centre would diverge.}
To ensure continuity of the shape profile or, in other words, to guarantee a closed surface, we impose the constraint that $r(\pi/2; \kappa_{\alpha}, {\change \gamma_{\alpha}})=r(\pi/2; \kappa_{\beta}, {\change \gamma_{\beta}})$ at the polar angle where, by construction, the two boundaries meet.
Having found the best set $\tilde{\kappa}_{\mu}, {\change \tilde{\gamma}_{\mu}}$, we calculate the acorn parameter
\begin{equation}
    \nu_{\mathrm{exp}}=\left({{\change \frac{1}{\tilde{\gamma}_{\alpha}}}+\frac{s}{\tilde{\kappa}_{\alpha}}}\right){\tilde{\kappa}_{\beta}} \,,
\end{equation}
where $s=1$ ($s=-1$) if the tissue--tissue contour is convex (concave),
and the base central angle
\begin{equation}
    \sigma_{\mathrm{exp}}=\cos^{-1}\left(-\frac{\tilde{\kappa}_{\beta}}{\change \tilde{\gamma}_{\beta}}\right)
\end{equation}
that characterise the epiblast.

Errors for the parameters are estimated by residual bootstrap. We generate a vector of residuals $\rho_{\mu} = \left(\delta r_{\mu 1}, \ldots, \delta r_{\mu N_{\mu}}\right)$, where $\delta r_{\mu i}=\mathcal{R}\left(\mathbf{r}^{\mathrm{exp}}_{\mu i}; \tilde{\kappa}_{\mu}, {\change \tilde{\gamma}_{\mu}} \right)$, for each boundary $\mu$. A residual bootstrap sample $\rho_{\mu}'=\left(\delta r'_{\mu 1}, \ldots, \delta r'_{\mu N_{\mu}}\right)$ is generated by randomly selecting $N_{\mu}$ elements from $\rho_{\mu}$ with replacement. Then, the cost function~\eqref{eq:costfunction} is evaluated at the modified datasets $\mathbf{r}'^{\,\mathrm{exp}}_{\mu}= \left\{\left( r\left(\theta_{\mu i}^{\mathrm{exp}}; \tilde{\kappa}_{\mu}, {\change \tilde{\gamma}_{\mu}}\right) +\delta r'_{\mu i},\ \theta_{\mu i}^{\mathrm{exp}},\ 0\right)\right\}_{i=1,\ldots,N_{\mu}} $ and minimised to obtain the corresponding best set of parameters. We repeat this process for 100 residual bootstrap samples to generate distributions of the fitting parameters, and use their standard deviation as uncertainties $\Delta\tilde{\kappa}_{\mu}, \Delta{\change \tilde{\gamma}_{\mu}}$. These give rise to
\begin{equation}
    \Delta\nu_{\mathrm{exp}} = \sqrt{\frac{\tilde{\kappa}_{\beta}^2}{\change \tilde{\gamma}_{\alpha}^4}{\change \Delta\tilde{\gamma}_{\alpha}^2} + \frac{\tilde{\kappa}_{\beta}^2}{\tilde{\kappa}_{\alpha}^4} \Delta\tilde{\kappa}_{\alpha}^2 + \frac{\nu_{\mathrm{exp}}^2}{\tilde{\kappa}_{\beta}^2} \Delta \tilde{\kappa}_{\beta}^2 }
\end{equation}
and
\begin{equation}
    \Delta\sigma_{\mathrm{exp}} =  \frac{ \sqrt{\dfrac{\tilde{\kappa}_{\beta}^2}{\change \tilde{\gamma}_{\beta}^4}{\change \Delta\tilde{\gamma}_{\beta}^2} + \dfrac{\Delta\tilde{\kappa}_{\beta}^2}{{\change \tilde{\gamma}_{\beta}^2}} }}{\sqrt{1-\dfrac{\tilde{\kappa}_{\beta}^2}{\change \tilde{\gamma}_{\beta}^2}}} \,.
\end{equation}

{\change

\subsection{Fitting the material parameters to the experimental alignment field}
The nematic alignment tensor captures the cell's orientation coherence, but not the polarity direction; its magnitude, however, can provide an unsigned estimation of the polarity magnitude. We define the experimental polarity magnitude $p_{\mathrm{exp}}\equiv \Lambda/\Lambda_{\mathrm{max}}$, where the strength of the nematic alignment $\Lambda(\mathbf{r})$ is normalised by its maximum value $\Lambda_{\mathrm{max}}$. 
In order to determine the model parameters, we fit the average epiblast shape (see \nameref{sec:shape_model_fitting}), and minimize the free energy functional~\eqref{eq:F} in this geometry for parameter ranges $\xi/R_0\in[1, 10^2]$, $\lambda/R_0\in[10^{-1}, 10^3]$ and $k_0/k_2\in[10^{-1}, 10^3]$ to obtain the corresponding theoretical field with magnitude $p$. In order to compare the experimental and theoretical results, we define a set of $N$ points $\mathbf{r}_i$ in the intersection of the theoretical and experimental embryo shapes. These points are chosen to be at a distance greater than $0.05$ from the symmetry axis and greater than $\ell$ from the closest boundary, where $\ell$ is given by the mean standard deviation of the experimentally determined average boundary. In this way, we exclude regions where the average experimental field has low statistics due to the different shapes of the individual embryos. Finally, we calculate the cost function 
\begin{equation}
    \mathcal{C}(\xi/R_0, \lambda/R_0, k_0/k_2)=\sum_{i=1}^N \left[p(\mathbf{r}_i;\xi/R_0, \lambda/R_0, k_0/k_2) - p_{\mathrm{exp}}(\mathbf{r}_i)\right]^2
\end{equation}
and identify the set of parameters where it attains its minimum value.

}




\section{Data availability}
The data that support the findings of this study will be made public upon publication.

\section{Code availability}
All the codes that support this study are available at \url{https://git.embl.de/guruciag/geometry-driven-defects}.



\section{Acknowledgements} 
We thank R.~Belousov, V.~Bondarenko, T.~Dullweber, J.~Ellenberg, J.~Elliott, I.~Estabrook, P.~Jentsch, L.~Manning, C.~Modes, T.~Quail, J.~Rombouts, and A.~Torres S\'anchez for useful discussions and valuable feedback on the manuscript. The Erzberger group is funded by the EMBL.
The Hiiragi laboratory was supported by the EMBL, and currently by the Hubrecht Institute, the European Research Council (ERC Advanced Grant ``SelforganisingEmbryo'' grant agreement 742732, ERC Advanced Grant ``COORDINATION'' grant agreement 101055287), Stichting LSH-TKI (LSHM21020), and Japan Society for the Promotion of Science (JSPS) KAKENHI grant numbers JP21H05038 and JP22H05166.
P.C.G. is supported by the EMBL Interdisciplinary Postdoctoral Fellowship (EIPOD4) programme under Marie Sk\l odowska-Curie Actions Cofund (grant agreement 847543), and by an Add-on Fellowship for Interdisciplinary Life Science of the Joachim Herz Stiftung. P.C.G. thanks Kyoto University for its hospitality during part of the preparation of this work. 

\section{Author contributions}
P.C.G., T.I., T.H. and A.E. designed the research. P.C.G. performed the numerical work. T.I. performed the experiments. T.I. and P.C.G. analysed the data and fitted the shape model. S.P. and P.C.G. calculated the experimental nematic field. P.C.G. prepared the figures with input from A.E. P.C.G. and A.E. wrote the paper with contributions from T.I., T.H. and S.P.

\section{Competing interests}
The authors declare no competing interests.


%


\clearpage
\setcounter{figure}{0}  
\renewcommand{\figurename}{Extended Data FIG.}

\begin{figure*}[htb]
    \centering
    \includegraphics[scale=0.85]{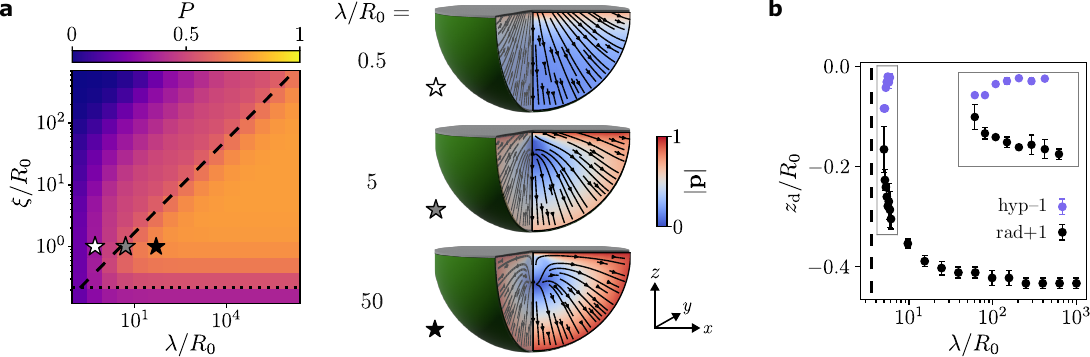}
    \caption{\textbf{The degree of global order and type of defects depend on the confining geometry.} 
    \textbf{a,}~Global degree of order $P$ [equation~\eqref{eq:globalOP}] of a hemispherical system with mixed $\mathbf p_0$. Order penetrates into the bulk more easily than in spherical confinement, obtaining $P>0.5$ in the limit $\lambda\to\infty$ for $\xi > 0.2R_0$ (dotted line). Stars denote the parameter values for example field configurations, where defects are created in order to accommodate the increasingly relevant boundary conditions ($\xi/ R_0=1$, $k_0/k_2 = 10$). The value $\zeta^*\approx 0.3R_0$ (dashed line) that separates defect-free and defect-containing configurations is independent of the confining geometry.
    \textbf{b,}~Increasing the anchoring length $\lambda$ changes the defect height $z_{\rm{d}}$ ($\xi/ R_0=1$, $k_0/k_2 = 10$).
    The hyp$-1$ bulk defect moves up and becomes a hyperbolic boojum on the surface, leaving behind only the rad$+1$ hedgehog. Points and error bars correspond to the mean and standard deviation of the position of the defect candidates (see~\nameref{sec:methods}). The vertical dashed line marks the value $\lambda^*=\xi^2/\zeta^*$.}
    \label{ext:hemisphere}
\end{figure*}

\begin{figure*}[htb]
    \centering
    \includegraphics[scale=0.85]{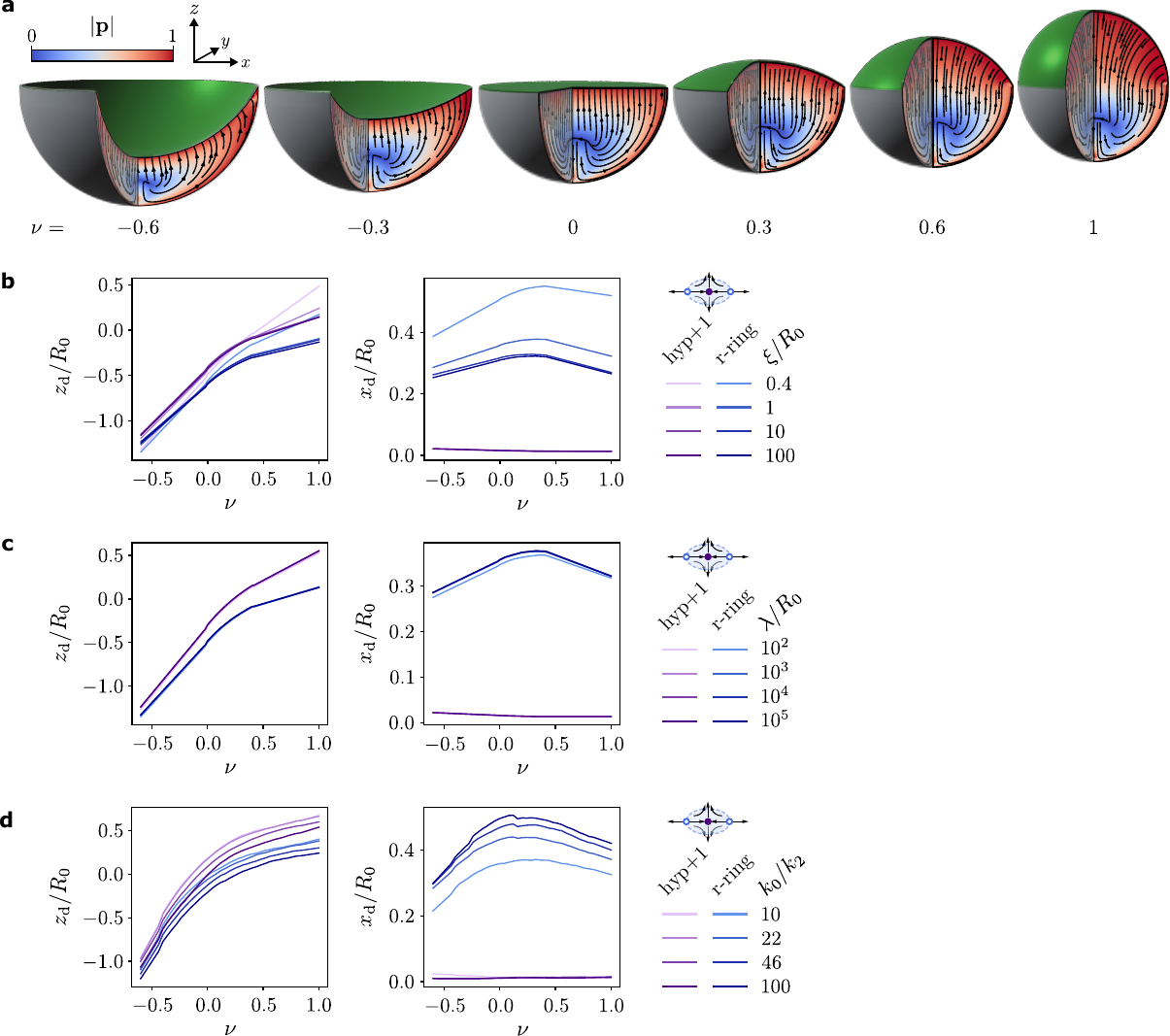}
    \caption{{\change \textbf{Flipping the mixed boundary preference gives rise to the combination of hyp$+1$ and r-ring for all geometries.} 
    \textbf{a,}~OP fields for different {\change values of the acorn parameter} $\nu$ at fixed material length scales $\lambda/ R_0=10^6$ and $\xi/ R_0=1$ with $k_0/k_2 = 10$.
    \textbf{b-d,}~Within the highly ordered regime, a hyp$+1$ point defect and a radial disclination ring are present in the bulk for the whole range of acorn parameter $\nu$. Their height $z_{\mathrm{d}}$ and distance to the symmetry axis $x_{\mathrm{d}}$ varies slightly with the correlation length $\xi$ (\textbf{b}) and the splay-to-bend ratio $k_0/k_2$ (\textbf{d}), but is unaffected by variations of the anchoring length $\lambda$ (\textbf{c}). Material parameters kept constant in each case are the same as in panel~\textbf{a}.
    }}
    \label{ext:flippedmixed}
\end{figure*}

\begin{figure*}[htb]
    \centering
    \includegraphics[scale=0.85]{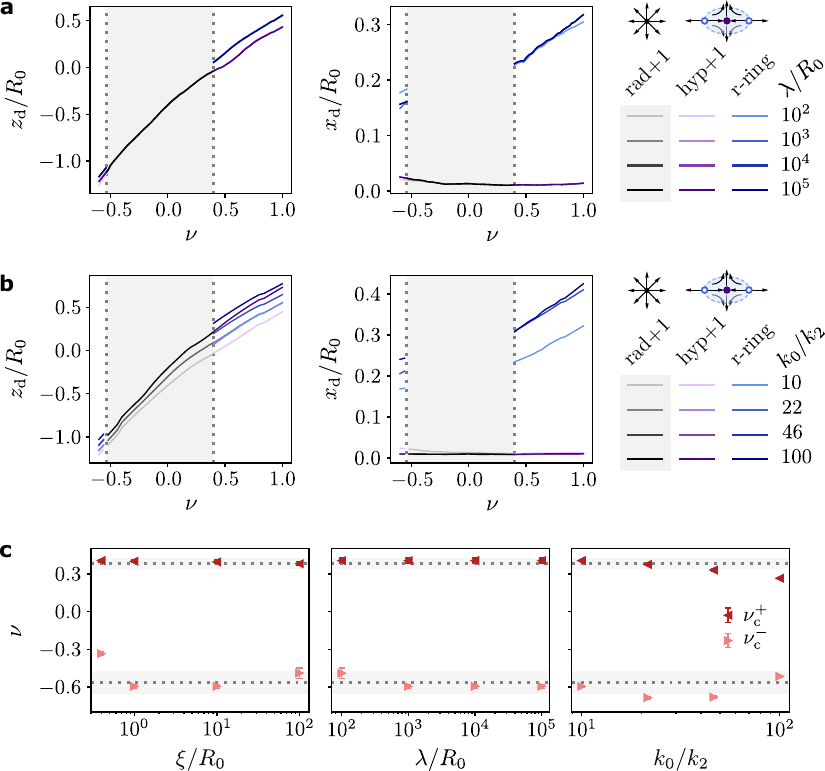}
    \caption{\textbf{The position of defects and the transitions between them are robust to parameter changes.} 
    \textbf{a,}~Within the highly ordered regime, the {\change height $z_{\mathrm{d}}$ and distance to the symmetry axis $x_{\mathrm{d}}$} of the defects is not affected by a three order of magnitude variation of the anchoring length ($\xi/R_0=1$, $k_0/k_2=10$).
    \textbf{b,}~{\change By increasing the splay-to-bend ratio $k_0/k_2$, i.e. by further enforcing the splay-dominated regime, the defect positions change only slightly ($\lambda/ R_0=10^6$, $\xi/ R_0=1$).}
    \textbf{c,}~The critical values of the acorn parameter where the transitions between defect configurations occur remain unchanged under variation of the mechanical parameters, giving $\nu_{\mathrm c}^+=\change{0.38\pm 0.04}$ and $\nu_{\mathrm c}^-=\change{-0.56\pm 0.09}$ (dotted lines and shaded area).
    }
    \label{ext:nucrit}
\end{figure*}

\begin{figure*}[htb]
    \centering
    \includegraphics[scale=0.85]{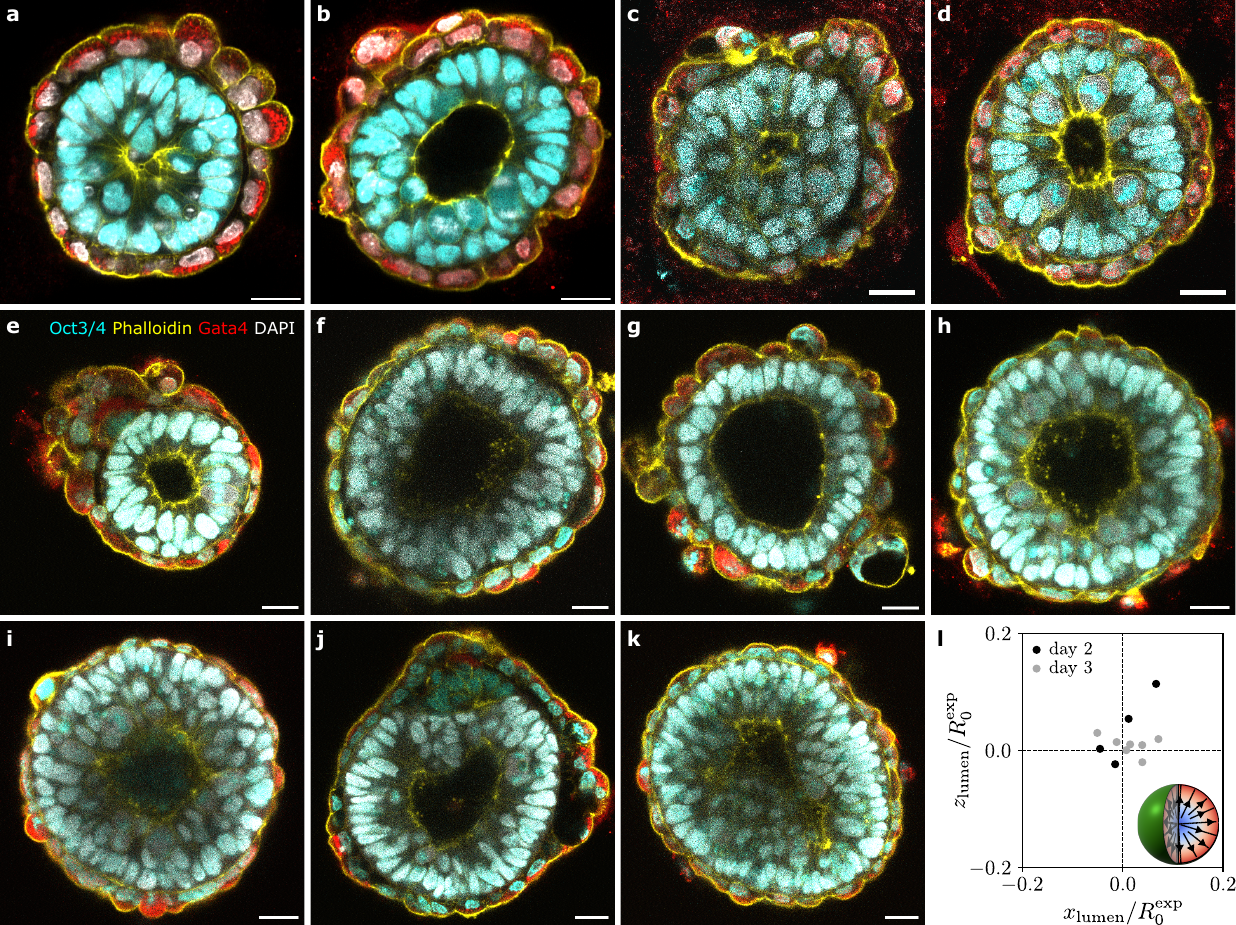}
    \caption{
    {\change 
    \textbf{ExE-devoid epiblasts form rosette patterns with a lumen in the centre.} 
    \textbf{a-k,}~Mouse blastocysts whose trophectoderm was removed, thus not forming ExE, were cultured for 2 days (\textbf{a-d}) or 3 days (\textbf{e-k}). Immunostaining for Oct3/4 (epiblast marker, cyan), phalloidin (membrane, yellow), Gata4 (visceral endoderm, red), and DAPI (nucleus for all cells, white). Scale bars \SI{20}{\micro\meter}.
    \textbf{l,}~Lumen centroid coordinates in the ExE-devoid embryos measured with respect to the corresponding epiblast centroids. A spherical system with homogeneous perpendicular anchoring (inset and Fig.~\ref{fig:defects_parameters}b) predicts lumen initiation in the centre (dashed lines). The volume $V_0^{\mathrm{exp}}$ of the epiblast tissue defines $R_0^{\mathrm{exp}}=\left(3V_0^{\mathrm{exp}}/4\pi\right)^{1/3}$ for each embryo.
    }
    }
    \label{ext:ExE-devoid}
\end{figure*}

\begin{figure*}[htb]
    \centering
    \includegraphics[scale=0.85]{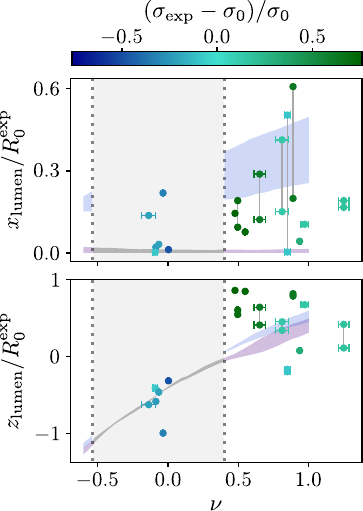}
    \caption{\textbf{Lumina vs. generalised-acorn parameters of the studied epiblasts.} 
    {\change Distance to the epiblast central axis $x_{\mathrm{lumen}}$ and height $z_{\mathrm{lumen}}$ of lumina} in terms of the individually fitted parameters of the generalised acorn model, $\nu_{\mathrm{exp}}$ and $\sigma_{\mathrm{exp}}$. The colour map represents the relative deviation from the ideal acorn shape, characterised by $\sigma_0=\pi/2$.
    Coloured shaded regions show the defect position ranges for ideal acorns with $\lambda/R_0=10^6$, $\xi/R_0\in[0.4,100]$ and $k_0/k_2 = 10$. The volume $V_0^{\mathrm{exp}}$ of the epiblast tissue defines $R_0^{\mathrm{exp}}=\left(3V_0^{\mathrm{exp}}/4\pi\right)^{1/3}$ for each embryo. Error bars are determined by residual bootstrapping of the epiblast boundaries (see \nameref{sec:methods}).
    }
    \label{ext:sigma}
\end{figure*}


\clearpage
\setcounter{figure}{0}  
\renewcommand{\figurename}{Supplementary Video}

\section{Captions of supplementary videos}

\begin{figure*}[htb]
    \centering
    \caption{\change \textbf{Cell dynamics, orientational organisation and initial lumen inflation in the mouse epiblast (time resolution 20 min).}
    An embryo expressing H2B-GFP and membrane-tdTomato was live-imaged with inverted light-sheet microscopy at $\SI{20}{\minute}$ time resolution from E4.5. Scale bars \SI{20}{\micro\meter}, time in hours:minutes.
    }
    \label{vid:epiblast_dynamics_20}
\end{figure*}

\begin{figure*}[htb]
    \centering
    \caption{\change \textbf{Cell dynamics, orientational organisation and initial lumen inflation in the mouse epiblast (time resolution 5 min).}
    An embryo expressing H2B-GFP and membrane-tdTomato was live-imaged with inverted light-sheet microscopy at $\SI{5}{\minute}$ time resolution from E5. Scale bars \SI{20}{\micro\meter}, time in hours:minutes.
    }
    \label{vid:epiblast_dynamics_5}
\end{figure*}

\begin{figure*}[htb]
    \centering
    \caption{\change \textbf{3D membrane segmentation of mouse epiblast tissue.} Embryos stained for Oct3/4 (epiblast marker, cyan) and phalloidin (actin, white) were processed with membrane segmentation. 2D $z$-stack slider and 3D volume visualization in Napari viewer.
    }
    \label{vid:3D_stack}
\end{figure*}


\clearpage

\section{Supplementary note}\label{sec:supinfo}
\setcounter{equation}{0}  
\setcounter{figure}{0}  
\renewcommand{\figurename}{Supplementary FIG.}


\subsection{{\change Minimisation of the free energy functional and d}erivation of the Euler-Lagrange equations}

In order to find the value of the vector field $\mathbf p(\mathbf r)$ that minimises the free energy functional
\begin{equation}\label{eq:F_supmat}
    \mathcal F[\mathbf p]
    =\int_{\Omega} \diff V f_{\mathrm B}(\mathbf p,\nabla\mathbf p) + \int_{\partial\Omega} \diff S f_{\mathrm S}(\mathbf p) \,,
\end{equation}
in a volume $\Omega$, we write the variation
\begin{equation}\label{eq:diff_F}
    \delta\mathcal{F}=\int_{\Omega}\diff V \left[ \frac{\partial f_{\mathrm{B}}}{\partial p_i}\delta p_i +\frac{\partial f_{\mathrm{B}}}{\partial \left(\nabla p_i\right)} \cdot\nabla\left(\delta p_i\right)\right]+
    \int_{\partial\Omega}\diff S\,\frac{\partial f_{\mathrm{S}}}{\partial p_i}\delta p_i \,,
\end{equation}
where Einstein summation convention over repeated indexes is used. 
The second term of the volume integral can be integrated by parts to yield
\begin{align}
    \int_{\Omega}\diff V \frac{\partial f_{\mathrm{B}}}{\partial \left(\nabla p_i\right)}\cdot\nabla\left(\delta p_i\right) &= \int_{\Omega}\diff V \Bigg\lbrace - \nabla\cdot\left[\frac{\partial f_{\mathrm{B}}}{\partial \left(\nabla p_i\right)}\right] \delta p_i +
    \nabla\cdot\left[\frac{\partial f_{\mathrm{B}}}{\partial \left(\nabla p_i\right)}\delta p_i \right]
    \Bigg\rbrace \nonumber\\
    &= -\int_{\Omega}\diff V\, \nabla\cdot\left[\frac{\partial f_{\mathrm{B}}}{\partial \left(\nabla p_i\right)}\right] \delta p_i + \int_{\partial\Omega}\diff S\, \hat{\mathbf n}\cdot \frac{\partial f_{\mathrm{B}}}{\partial \left(\nabla p_i\right)} \delta p_i
\end{align}
with $\hat{\mathbf{n}}$ the outward normal.
Thus, equation~\eqref{eq:diff_F} becomes
\begin{equation}
    \delta\mathcal{F}=\int_{\Omega}\diff V \left[ \frac{\partial f_{\mathrm{B}}}{\partial p_i} -\nabla\cdot\frac{\partial f_{\mathrm{B}}}{\partial \left(\nabla p_i\right)}\right] \delta p_i+
    \int_{\partial\Omega}\diff S \left[\frac{\partial f_{\mathrm{S}}}{\partial p_i}+ \hat{\mathbf n}\cdot \frac{\partial f_{\mathrm{B}}}{\partial \left(\nabla p_i\right)}\right]\delta p_i \,.
\end{equation}
If the minimising field $\mathbf{p}$ were to have a fixed value on the boundary (essential boundary conditions), $\delta p_i \big|_{\partial\Omega}=0$ and the surface integral would vanish directly. In our case, however, we take boundary conditions on $\mathbf{p}$ to be natural, meaning that they are not imposed externally, but are \textit{naturally} derived form the problem and satisfied after a solution has been found. Hence, for arbitrary variations $\delta p_i$, we enforce
\begin{equation}
    \frac{\partial f_{\mathrm S}}{\partial p_i}+\hat{\mathbf n}\cdot\frac{\partial f_{\mathrm B}}{\partial \left(\nabla p_i\right)}=0 \quad\text{in }\partial\Omega \,.
\end{equation}
These equations ensure continuity of $\mathbf p$ on the surface and leave us only with the volume integral. Finally, asking that $\delta\mathcal{F}=0$ gives rise to the set of coupled partial differential equations
\begin{equation}
    \frac{\partial f_{\mathrm B}}{\partial p_i}-
    \nabla\cdot\frac{\partial f_{\mathrm B}}{\partial\left( \nabla p_i\right)} =0 \quad\text{in } \Omega \,,
\end{equation}
which are the Euler-Lagrange equations of the problem.

{\change
\subsection{Relation between the elastic contribution $f_{\mathrm{E}}$ and the Frank free energy density}

The expression used in this work for the elastic energy density is
\begin{equation}\label{eq:fE_SN}
    f_{\mathrm E} =\frac{k_0}{2}\left(\nabla\cdot\mathbf p\right)^2 + \frac{k_1}{2}{\left[\hat{\mathbf p}\cdot \left(\nabla\times\mathbf p\right)\right]^2} + \frac{k_2}{2}{\left[\hat{\mathbf p}\times \left(\nabla\times\mathbf p\right)\right]^2} \,.
\end{equation}
In this section we discuss its relation to the more standard Frank free energy density.

\subsubsection{In the one-constant approximation}

Typically~\cite{de1993physics,alert2019active,pallares2023stiffness,ibrahimi2023deforming}, the elastic energy density of a vector field $\mathbf{p}$ is taken to be 
\begin{equation}\label{eq:Frank_dipjdipj}
    f_{\mathrm{Frank}}=\frac{k}{2}(\partial_i p_j)(\partial_i p_j)
\end{equation}
with $\partial_i\equiv\partial/\partial x_i$ and $k>0$ the Frank constant. Using explicitly that $\mathbf{p}=|\mathbf p|\hat{\mathbf{p}}$ and $\hat p_j\partial_i \hat p_j\propto\partial_i(\hat p_j\hat p_j)=\partial_i(1) =0$, it can be rewritten as
\begin{equation}
    f_{\mathrm{Frank}} =\frac{k}{2}\left[\left(\nabla|\mathbf p|\right)^2 + |\mathbf p|^2(\partial_i \hat p_j)(\partial_i \hat p_j)\right] \,.
\end{equation}
The last term can be decomposed into four terms corresponding to the splay, twist, bend and saddle-splay modes~\cite{selinger2018interpretation}, thus giving rise to
\begin{equation}\label{eq:Frank_hatp_expansion}
    f_{\mathrm{Frank}} = 
    \frac{k}{2}\left\{ \left(\nabla|\mathbf p|\right)^2 + |\mathbf p|^2\left[
    \left(\nabla\cdot\hat{\mathbf p}\right)^2 + \left[\hat{\mathbf p}\cdot \left(\nabla\times\hat{\mathbf p}\right)\right]^2 + \left[\hat{\mathbf p}\times \left(\nabla\times\hat{\mathbf p}\right)\right]^2\right]\right\} + f_{\mathrm{saddle-splay}}
\end{equation}
with $f_{\mathrm{saddle-splay}}= - k|\mathbf p|^2\nabla\cdot\left[ \hat{\mathbf p}(\nabla\cdot\hat{\mathbf p}) + \hat{\mathbf p}\times(\nabla\times\hat{\mathbf p}) \right]/2$. 

Let us now consider the elastic energy density~\eqref{eq:fE_SN} with $k_0=k_1=k_2\equiv k$, that is,
\begin{equation}\label{eq:fEp}
    f_{\mathrm E'} =\frac{k}{2}\left\{\left(\nabla\cdot\mathbf p\right)^2 + {\left[\hat{\mathbf p}\cdot \left(\nabla\times\mathbf p\right)\right]^2} + {\left[\hat{\mathbf p}\times \left(\nabla\times\mathbf p\right)\right]^2}\right\} \,.
\end{equation}
Notice that the prime is used throughout this section to emphasise the one-constant approximation.
We can expand each term as
\begin{align}
    \left(\nabla\cdot\mathbf p\right)^2 &
    = |\mathbf p|^2\left(\nabla\cdot\hat{\mathbf p} \right)^2 + (\nabla|\mathbf p|\cdot\hat{\mathbf p})^2 + 2|\mathbf p|(\nabla|\mathbf p|\cdot \hat{\mathbf p})(\nabla\cdot\hat{\mathbf p})  \label{eq:decomp_splay}\\
    \left[\hat{\mathbf p}\cdot \left(\nabla\times{\mathbf p}\right)\right]^2 &
    = |\mathbf p|^2\left[\hat{\mathbf p}\cdot(\nabla\times\hat{\mathbf p})\right]^2 \label{eq:decomp_twist}\\
    \left[\hat{\mathbf p}\times \left(\nabla\times{\mathbf p}\right)\right]^2 &
    = |\mathbf p|^2\left[\hat{\mathbf p}\times(\nabla\times\hat{\mathbf p})\right]^2 + \left[\hat{\mathbf p}\times(\nabla|\mathbf p|\times\hat{\mathbf p})\right]^2 + 2|\mathbf p|\left[ \hat{\mathbf p}\times (\nabla|\mathbf p|\times\hat{\mathbf p}) \right]\cdot\left[\hat{\mathbf p}\times(\nabla\times\hat{\mathbf p})\right] \label{eq:decomp_bend}
\end{align}
The first terms of each identity are equal to the the terms in square brackets in equation~\eqref{eq:Frank_hatp_expansion}, and the sum of the second terms of identities~\eqref{eq:decomp_splay} and~\eqref{eq:decomp_bend} is equal to $\left(\nabla|\mathbf p|\right)^2$. The sum of their third terms, however, is not present in equation~\eqref{eq:Frank_hatp_expansion}; we use it to write the cross-term energy density
\begin{equation}
    f_{\mathrm{cross'}}= 
    k \left\{({\mathbf p}\cdot \nabla|\mathbf p|)(\nabla\cdot\hat{\mathbf p}) + \left[ {\mathbf p}\times (\nabla|\mathbf p|\times\hat{\mathbf p}) \right]\cdot\left[\hat{\mathbf p}\times(\nabla\times\hat{\mathbf p})\right]\right\} \,.
\end{equation}
With this, we have that $f_{\mathrm E'}= f_{\mathrm{Frank}} - f_{\mathrm{saddle-splay}} + f_{\mathrm{cross'}}$.

We now evaluate numerically the difference between the standard Frank free energy density~\eqref{eq:Frank_dipjdipj} and the approximated expression~\eqref{eq:fEp}. 
\begin{figure*}[htb]
    \centering
    \includegraphics[scale=0.85]{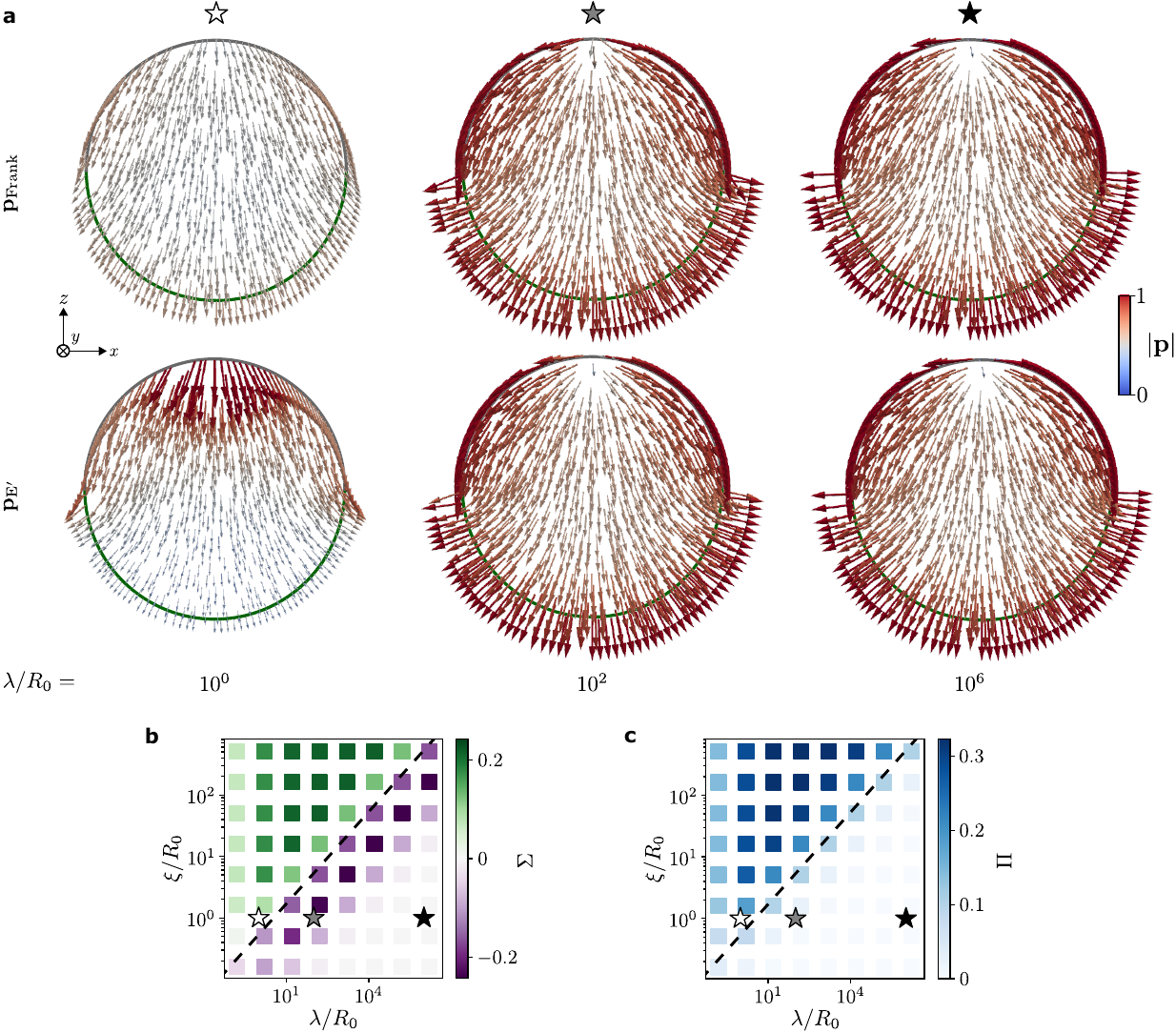}
    \caption{
    \textbf{Numerical evaluation of the difference between the standard Frank free energy density and the elastic energy terms in the one-constant approximation.} 
    \textbf{a,}~Central slices of the 3D OP-field configurations obtained using $f_{\mathrm{Frank}}$ and $f_{\mathrm E'}$ in a sphere for $\xi/R_0=1$ and different values of $\lambda/R_0=10$. Colour map and arrow size encode the magnitude of the OP.
    \textbf{b-c,}~Relative energy difference [\textbf{b}, equation~\eqref{eq:Ediff}] and average OP difference [\textbf{c}, equation~\eqref{eq:pdiff}] vs. $\lambda/R_0$ and $\xi/R_0$ in spherical confinement. Both quantities indicate an excellent agreement of the results for small values of $\zeta$. 
    As a reference, the dashed line shows the value $\zeta^*\approx0.3R_0$ of the main text. Stars highlight the sets of parameters where the configurations of panel~\textbf{a} were obtained.
    }
    \label{sup:FrankEp}
\end{figure*}
We consider the free energy functional given by equation~\eqref{eq:F_supmat}, and use $f_{\mathrm B}=f_{\mathrm A}+f_{\eta}$ with $f_{\mathrm A}$ as defined in the main text and $f_{\eta}$ a placeholder for $f_{\mathrm{Frank}}$ or $f_{\mathrm{E'}}$. The surface energy density $f_{\mathrm S}$ is also taken to be the same as in the main text, with a mixed preferred orientation in the boundary. We minimise equation~\eqref{eq:F_supmat} in a sphere without assuming any symmetry to find the corresponding OP fields $\mathbf{p}_{\eta}$ (Supplementary Fig.~\ref{sup:FrankEp}a). In order to evaluate discrepancies in the results quantitatively, we define the relative energy difference
\begin{equation}\label{eq:Ediff}
    \Sigma\equiv\frac{H_{\mathrm{Frank}}-H_{\mathrm{E'}}}{H_{\mathrm{Frank}}} \,,
\end{equation}
where $H_{\eta}$ is the total energy of the system with OP $\mathbf{p}_{\eta}$ and elastic energy density $f_{\eta}$, and the average difference between the OP fields
\begin{equation}\label{eq:pdiff}
    \Pi\equiv\frac{1}{N_{\mathrm{s}}}\sum_{i=1}^{N_{\mathrm{s}}} \left|\mathbf{p}_{\mathrm{Frank}}(\mathbf{r}_i)-\mathbf{p}_{\mathrm{E'}}(\mathbf{r}_i)\right| \,,
\end{equation}
where $N_{\mathrm{s}}$ is the number of nodes in the mesh.
Both $\Sigma$ and $\Pi$ show that $\mathbf{p}_{\mathrm{Frank}}$ and $\mathbf{p}_{\mathrm{E'}}$ present excellent agreement in the low $\zeta=\xi^2/\lambda$ regime (Supplementary Figs.~\ref{sup:FrankEp}b-c). Note that both fields present a radial boojum on the top cap like that of Fig.~1c (bottom line, second leftmost panel) in the main text.
Conversely, differences between $\mathbf{p}_{\mathrm{Frank}}$ and $\mathbf{p}_{\mathrm{E'}}$ are most significant in the high $\zeta$ regime. Nonetheless, both OP fields are topologically equivalent to each other and to the uniform field.

\subsubsection{Leaving the one-constant approximation}

We now focus on the elastic energy density of equation~\eqref{eq:fE_SN} in the splay-dominated regime ($k_0\gg k_2$). Minimisation of equation~\eqref{eq:F_supmat} in the full 3D, symmetry-agnostic case leads to an OP field consisting on a positive hyperbolic hedgehog on the $z$ axis and a radial disclination ring around it (Supplementary Fig.~\ref{sup:3Dring}a).
This supports the assumption of axial symmetry made in the main text (Supplementary Fig.~\ref{sup:3Dring}b), and rules out the possibility of spontaneous selection of a given value of the azimuth angle $\phi$ for an off-axis point defect instead of the ring.
\begin{figure*}[hbt]
    \centering
    \includegraphics[scale=0.85]{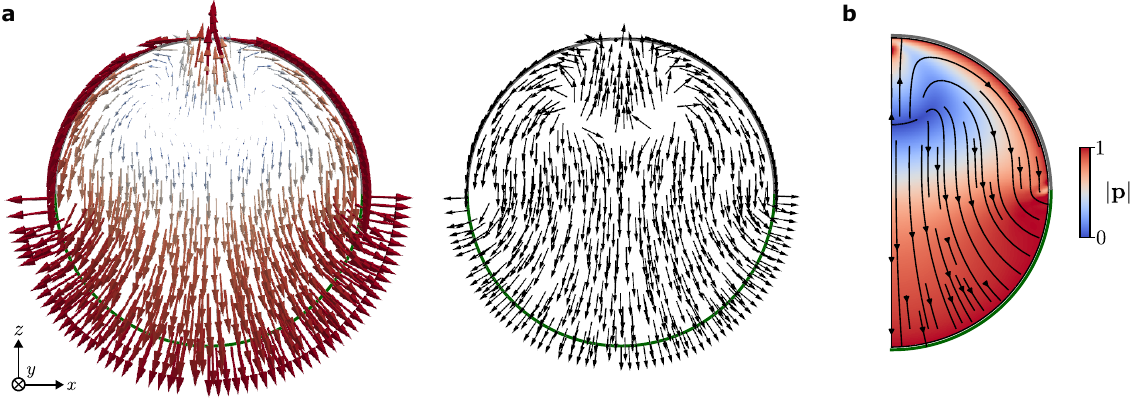}
    \caption{
    \textbf{3D solution consists of a positive hyperbolic hedgehog and a radial disclination ring.} 
    \textbf{a,}~Central slice of the 3D OP field obtained with $f_{\mathrm E}$ in the splay-dominated regime and spherical confinement ($\lambda/R_0=10^6$, $\xi/R_0=1$, $k_0/k_2=10$). Colour map and arrow size encode the magnitude of the OP (left). For clarity, a normalised, rescaled version of the OP is also shown (right). 
    \textbf{b,} The same configuration is obtained in the main work when the assumption of axial symmetry is made. Streamlines and colour map depict the field direction and magnitude, respectively.
    }
    \label{sup:3Dring}
\end{figure*}

In this more general case, the cross-term energy density is given by
\begin{equation}
    f_{\mathrm{cross}} = k_0 ({\mathbf p}\cdot \nabla|\mathbf p|)(\nabla\cdot\hat{\mathbf p}) + k_2\left[ {\mathbf p}\times (\nabla|\mathbf p|\times\hat{\mathbf p}) \right]\cdot\left[\hat{\mathbf p}\times(\nabla\times\hat{\mathbf p})\right] \,.
\end{equation}
We use this expression to define the error score
\begin{equation}\label{eq:errscore}
    \epsilon\equiv
    \frac{H_{\mathrm{cross}}}{H_{\mathrm{E}}} \,,
\end{equation}
where $H_{\mathrm{cross}}=\int_{\Omega}\diff V f_{\mathrm{cross}}$.
This value is a proxy for the previous relative energy difference and average OP difference, which allows us to approximately evaluate the discrepancy between the numerical results using equation~\eqref{eq:fE_SN} and the corresponding Frank free energy density, without having to compute the latter. Supplementary Fig.~\ref{sup:errorscore} present the error score of the results depicted in the main text.
\begin{figure*}[htb]
    \centering
    \includegraphics[scale=0.85]{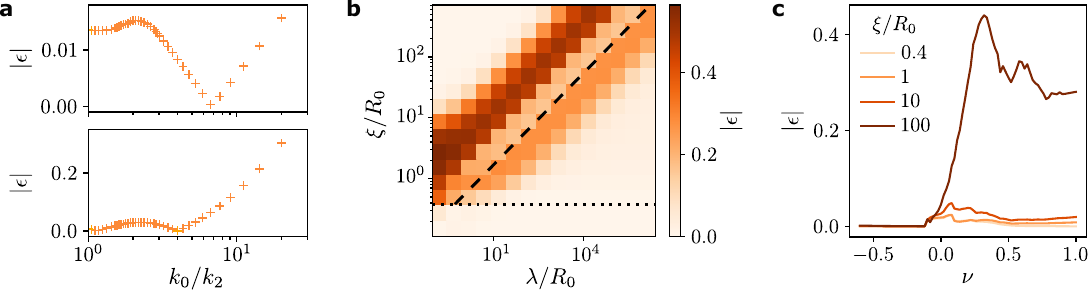}
    \caption{
    \textbf{Error score [equation~\eqref{eq:errscore}] for the results in the main text.} 
    \textbf{a,}~Error score as a function of $k_0/k_2$ for longitudinal parallel (upper panel) and mixed (lower panel) preferred orientations in spherical confinement ($\lambda/R_0=100$, $\xi/R_0=1$).
    \textbf{b,}~Error score vs. $\lambda/R_0$ and $\xi/R_0$ in spherical confinement with mixed $\mathbf{p}_0$ in the splay-dominated regime ($k_0/k_2=10$).
    \textbf{c,}~Error score as a function of the acorn parameter $\nu$ for different values of $\xi/R_0$ ($\lambda/R_0=10^6$, $k_0/k_2=10$).
    }
    \label{sup:errorscore}
\end{figure*}

}


\end{document}